\documentclass[journal,10pt]{IEEEtran}

\usepackage{amsmath,amssymb}
\usepackage{subfigure}
\usepackage{graphicx,graphics,color,psfrag}
\usepackage{cite,balance}
\usepackage{algorithm}
\usepackage{accents}
\usepackage{amsthm}
\usepackage{bm}
\usepackage{url}
\usepackage{algorithmic}
\usepackage[english]{babel}
\usepackage{multirow}
\usepackage{enumerate}
\usepackage{cases}
\usepackage{stfloats}
\usepackage{dsfont}
\usepackage{color,soul}
\usepackage{amsfonts}
\usepackage{tcolorbox}

\usepackage{cite,graphicx,amsmath,amssymb}
\usepackage{fancyhdr}
\usepackage{hhline}
\usepackage{graphicx,graphics}
\usepackage{array,color}
\usepackage{amsmath}
\usepackage{amsthm}
\usepackage[flushleft]{threeparttable}
\IEEEoverridecommandlockouts

\newtheorem{proposition}{Proposition}
\newtheorem{remark}{Remark}

\include{header}

\newcommand{\mv}[1]{\mbox{\boldmath{$ #1 $}}}

\makeatletter
\def\endthebibliography{%
	\def\@noitemerr{\@latex@warning{Empty `thebibliography' environment}}%
	\endlist
}
\makeatother

\begin{document}

%
%
%
	\title{Optimal Transmit Beamforming for Integrated Sensing and Communication}
\author{Haocheng Hua, \textit{Student Member, IEEE}, Jie Xu, \textit{Senior Member, IEEE}, and Tony Xiao Han, \textit{Senior Member, IEEE} \\
	\thanks{Copyright (c) 2015 IEEE. Personal use of this material is permitted. However, permission to use this material for any other purposes must be obtained from the IEEE by sending a request to pubs-permissions@ieee.org.}
	\thanks{Manuscript received October 7, 2022; revised January 19, 2023; accepted March 23, 2023. 
	Part of this paper has been presented in IEEE Global Communications Conference (GLOBECOM), Madrid, Spain, December 7-11, 2021 \cite{hua2021transmit}.
	The associate editor coordinating the review of this article and approving it for publication was Zilong Liu. \textit{(Corresponding author: Jie Xu.)}
} 	
	\thanks{H. Hua and J. Xu are with the School of Science and Engineering (SSE) and the Future Network of Intelligence Institute (FNii), The Chinese University of Hong Kong (Shenzhen), Shenzhen, China (e-mail: haochenghua@link.cuhk.edu.cn, xujie@cuhk.edu.cn).}
	\thanks{T. X. Han is with the 2012 lab, Huawei, Shenzhen 518129, China (e-mail: tony.hanxiao@huawei.com). }
}

\vspace{-0.5cm}
\markboth{}{}
\maketitle


%

\maketitle

\begin{abstract}
	This paper studies the transmit beamforming in a downlink integrated sensing and communication (ISAC) system, where a base station (BS) equipped with a uniform linear array (ULA) sends combined information-bearing and dedicated radar signals to perform downlink multiuser communication and radar target sensing simultaneously. We consider two radar sensing design criteria, including the conventional sensing beampattern matching and the newly proposed minimum weighted beampattern gain maximization, respectively. Under this setup, we maximize the radar sensing performance while ensuring the communication users' individual signal-to-interference-plus-noise ratio (SINR) requirements subject to the BS's maximum transmit power constraints.
	In particular, we consider two types of communication receivers, namely Type-I and Type-II receivers, which do not have and do have the capability of cancelling the interference from the {\emph{a-priori}} known dedicated radar signals, respectively. Under both Type-I and Type-II receivers, the beampattern matching and minimum weighted beampattern gain maximization problems are non-convex and thus difficult to be optimally solved in general. Fortunately, via applying the semidefinite relaxation (SDR) technique, we obtain the globally optimal solutions to these problems by rigorously proving the tightness of their SDRs. Besides, for both design criteria, we show that dedicated radar signals are generally beneficial in enhancing the system performance with both types of receivers under general channel conditions, while the dedicated radar signals are not required with Type-I receivers under the special case with line-of-sight (LOS) communication channels.
	Numerical results show that the minimum weighted beampattern gain maximization design significantly outperforms the beampattern matching design, in terms of much higher beampattern gains at the worst-case sensing angles and much lower computational complexity in solving the corresponding problems. It is also shown that by exploiting the capability of cancelling the interference caused by the dedicated radar signals, the case with Type-II receivers results in better sensing performance than that with Type-I receivers and other conventional designs. 
\end{abstract}



\IEEEpeerreviewmaketitle


%
%

\section{Introduction}\label{sec:intro}

Future wireless networks are envisaged to support emerging applications such as autonomous driving in vehicle-to-everything (V2X), smart traffic control, virtual/augmented reality (AR/VR), smart home, unmanned aerial vehicles (UAVs), and factory automation \cite{rahman2020enabling}, which require ultra reliable and low-latency sensing,  communication, and computation. This thus calls for a paradigm shift in wireless networks, from the conventional communications only design, to a new design with sensing-communication-computation integration. Towards this end, integrated sensing and communication (ISAC)\footnote{In the literature, the ISAC is also referred to as joint radar communications (JRC) \cite{liu2020joint}, joint communication and radar (JCR) \cite{kumari2019adaptive}, dual-functional radar communications (DFRC) \cite{hassanien2016signaling,liu2018toward}, radar-communications (RadCom) \cite{sturm2011waveform}, and joint communication and radar/radio sensing (JCAS) \cite{rahman2020enabling}.}  \cite{liu2022integrated, liu2020joint,kumari2019adaptive,sturm2009ofdm,hassanien2016signaling,sturm2011waveform,donnet2006combining,dokhanchi2019mmwave,liu2018toward,liu2018mu,Eldar2020joint,mccormick2017simultaneous_I,mccormick2017simultaneous_II,liyanaarachchi2021joint, zhang2022holographic, wang2022noma, xu2021rate} has recently attracted growing research
interests in both academia and industry, in which the radar sensing capabilities are integrated into wireless networks (e.g., beyond-fifth-generation (B5G)/sixth-generation (6G) cellular \cite{wild2021joint} and WiFi 7 \cite{noauthor_undated-pr,tan2020wi}) for a joint communication and sensing design.
On one hand, with such integration, ISAC can significantly enhance the spectrum utilization efficiency and cost efficiency by exploiting the dual use of radio signals and infrastructures for both radar sensing and wireless communication \cite{wild2021joint}.
On the other hand, the recent advancements of millimeter wave (mmWave)/terahertz (THz), wideband transmission (on the order of 1 GHz), and massive antennas in wireless communications can be utilized in radar sensing to provide high resolution (in both range and angular) and accuracy (in detection and estimation), thus meeting the stringent sensing requirements of emerging applications. In the industry, Huawei and Nokia envisioned ISAC as one of the key technologies for 6G \cite{wild2021joint,tan2021integrated}, and IEEE 802.11 formed the WLAN Sensing Task Group IEEE 802.11bf in September 2020, with the objective of incorporating wireless sensing as a new feature for next-generation WiFi systems (e.g., Wi-Fi 7) \cite{noauthor_undated-pr,tan2020wi}.

While ISAC is a relatively new research topic that emerged recently in the communications society,  the interplay between wireless communications and radar sensing has been extensively investigated in the literature, some examples of which include the communication and radar spectrum sharing (CRSS) \cite{zheng2019radar,labib2017coexistence,sodagari2012projection, mahal2017spectral, saruthirathanaworakun2012opportunistic,  rihan2018optimum,liu2018mimo,qian2018joint,li2016optimum, liu2018mu,liu2017robust, chiriyath2017radar} and the radar-centric communication integration \cite{li2019integrated,sahin2017filter,zhang2017modified,zheng2019radar,hassanien2016signaling,huang2020majorcom,wu2020waveform,wang2018dual,ma2018novel,hassanien2016dual,hassanien2016phase}. In particular, CRSS features the spectrum sharing between separate radar sensing and communication systems, for which a large amount of research efforts have been put on managing the inter-system interference via, e.g.,  interference null-space projection \cite{sodagari2012projection, mahal2017spectral}, opportunistic spectrum sharing \cite{saruthirathanaworakun2012opportunistic}, and transmit beamforming optimization \cite{rihan2018optimum,liu2018mimo,qian2018joint,li2016optimum,liu2018mu,liu2017robust}. However, CRSS generally has limited performance in sensing and communication due to the limited coordination and information exchange between them \cite{rahman2020enabling}.
In another line of research called radar-centric communication integration, the radar signals are revised to convey information, thus incorporating communication capabilities in conventional radar systems. For instance, the authors in \cite{li2019integrated,sahin2017filter,zhang2017modified} incorporated continuous phase modulation (CPM) into the conventional linear frequency modulation (LFM) radar waveform to
convey information, \cite{hassanien2016phase,hassanien2016dual} embedded the information bits via controlling the amplitude and phase of radar side-lobes and waveforms, and \cite{huang2020majorcom,wu2020waveform,wang2018dual,ma2018novel} modified the radar waveforms via index modulation to represent information by the indexes of antennas, frequencies, or codes instead of directly modifying the radar waveforms.
However, the radar-centric communication integration may lead to low communication data rates as the radar signaling is highly suboptimal for data transmission in general.

Different from the CRSS and radar-centric communication integration, the ISAC shares the same wireless infrastructures for radar sensing and communication, and jointly optimizes the transceivers design for both tasks.
In particular, thanks to its great success in wireless communications \cite{goldsmith2005wireless, heath2018foundations} and radar sensing \cite{fishler2004mimo, li2007mimo}, the multi-antenna or multiple-input multiple-output (MIMO) technique is expected to play an important role in ISAC, where the spatial degrees of freedom (DoF) can be exploited via transmit beamforming for simultaneous multi-target sensing and multi-user communication (see, e.g., \cite{mccormick2017simultaneous_I,mccormick2017simultaneous_II,liu2018toward,liu2018mu,Eldar2020joint,liyanaarachchi2021joint, zhang2022holographic, wang2022noma, xu2021rate}). Such benefit may become even more considerable in B5G/6G, due to the employment of massive MIMO \cite{larsson2014massive} and mmWave/THz \cite{heath2016overview}. How to design the transmit signal waveforms and the associated transmit beamforming is an essential issue to be dealt with in ISAC with multiple antennas. In the literature, initial work \cite{mccormick2017simultaneous_I,mccormick2017simultaneous_II} considered the simplified case with one single communication user and one single target, which were then extended to the case with multiple receivers and multiple sensing targets \cite{liu2018toward,liu2018mu,Eldar2020joint,liyanaarachchi2021joint, xu2021rate, wang2022noma,zhang2022holographic}. For instance, the authors in \cite{liu2018toward} optimized the transmit beamformers to minimize the multiuser interference at communication users, while matching a desired transmit beampattern for sensing. The authors in \cite{liu2018mu,Eldar2020joint} then considered the beampattern matching problem for sensing, while ensuring the signal-to-interference-plus-noise ratio (SINR) constraints at communication receivers subject to per-antenna transmit power constraints at the base station (BS). 
In particular, \cite{liu2018mu} considered that the information signals are directly reused for both communication and sensing, in which the generally sub-optimal solutions to the SINR-constrained beampattern matching problems are obtained by the technique of semidefinite relaxation (SDR). The design of reusing information signals, however, may lead to limited DoF for sensing, especially when the number of users is smaller than that of transmit antennas. To resolve this issue, \cite{Eldar2020joint} proposed to use dedicated radar signals combined with information signals to provide full DoF for sensing, based on which the resultant SINR-constrained beampattern matching problems are optimally solved. 
More recently, to reduce the implementation complexity and cost of fully digital MIMO, \cite{liyanaarachchi2021joint} and \cite{zhang2022holographic} studied the hybrid beamforming for ISAC in full-duplex systems and with holographic MIMO, respectively. Moreover, advanced multiple access techniques such as non-orthogonal multiple access (NOMA) \cite{wang2022noma} and rate-splitting multiple access (RSMA) \cite{xu2021rate} were exploited for enhancing ISAC performance.



Despite the above progress, these prior works still have the following limitations. First, although dedicated radar signals provide additional spatial DoF for sensing, they also introduce harmful interference towards the communication users, which may lead to compromised ISAC performance. Notice that the dedicated radar signals can be set as {\it a-priori} known pseudorandom sequences \cite{richards2005fundamentals}, which can thus be known prior to the transmission at both the transmitter and communication receivers. This thus provides a new opportunity to exploit the interference cancellation for enhancing the ISAC performance, which, however, has not been investigated in the literature yet.
Next, the SINR-constrained beampattern matching has been widely adopted as the ISAC design criterion, which normally leads to quadratic semidefinite programs (QSDPs), incurring high computational complexity in finding the optimal solution in general. It is thus desirable to investigate new ISAC design criterion and solution approaches with enhanced performance but lower complexity. These issues motivate the current work.

This paper studies the transmit beamforming design in a downlink ISAC system, in which a single BS equipped with a uniform linear array (ULA) sends combined information-bearing and dedicated radar signals to perform downlink multiuser communication and multi-target radar sensing simultaneously.
Our main results are listed as follows.
\begin{itemize}
	\item {\bf New communication receivers design for ISAC}: To exploit the property that radar signals are {\it a-priori} known, we present a new type of communication receiver design, which has the capability of cancelling the interference caused by  radar signals prior to decoding the information signals. To differentiate from the conventional communication receiver, we refer to Type-I and Type-II receivers as those without and with such radar interference cancellation capabilities, respectively. Accordingly, we investigate the joint transmit beamforming for ISAC with the two types of receivers.
	\item {\bf New ISAC design criterion}: Besides the conventional beampattern matching design criterion, we consider a new design criterion for sensing, which aims to maximize the minimum weighted  beampattern gain at the sensing angles of interest. Accordingly, by considering each type of communication receivers, we formulate two ISAC design problems, with the conventional objective of minimizing the beampattern matching error and the new objective of maximizing the minimum weighted  beampattern gain, respectively, subject to the minimum SINR constraints at individual communication receivers. The newly considered minimum weighted  beampattern gain maximization design will be shown to result in both better sensing beampatterns and lower solution complexity, than the conventional beampattern matching design.
	\item {\bf Optimal joint transmit beamforming solutions}: With both Type-I and Type-II receivers, the formulated beampattern matching and minimum weighted  beampattern gain maximization problems are non-convex and thus difficult to be optimally solved in general. Fortunately, by applying the SDR technique, we obtain their globally optimal solutions by rigorously proving that the SDRs are always tight for our considered four problems.
	\item {\bf Insights in joint transmit beamforming}: We rigorously prove that adding dedicated radar signals is generally beneficial in improving the ISAC performance for all the four designs with different design criteria and different types of receivers, by exploiting the full spatial DoF. In particular, with Type-II receivers, using dedicated radar signals is shown to always yield better sensing performance at the interested angles; while with Type-I receivers, we rigorously prove that there is no need to add dedicated radar signals in the special case with line-of-sight (LOS) communication channels.
	\item {\bf  Numerical results}: Finally, we provide numerical results to validate the performance of the proposed designs. It is shown that the newly proposed minimum weighted  beampattern gain maximization design yields enhanced  sensing performance at the interested angles, and also enjoys significantly lower computational complexity in solving the corresponding design problems in terms of execution time, as compared to the conventional  beampattern matching counterpart. It is also shown that under both design criteria, the case with Type-II receivers leads to significantly enhanced radar sensing performance as compared to that with Type-I receivers and other conventional designs.
\end{itemize}



The remainder of this paper is organized as follows. Section \ref{sec:system1} presents the system model. Section \ref{Section-joint} and \ref{S-maxmin} present the transmit beamforming designs under SINR-constrained beampattern matching design criterion and minimum weighted beampattern gain maximization design criterion, respectively. Section \ref{Section_results} provides numerical results to validate the performance of our proposed designs. Section \ref{Section_conclusion} concludes this paper.

{\it Notations:} Boldface letters refer to vectors (lower  case) or matrices (upper case). For a square matrix $\bm{S}$, ${\text{tr}}(\bm{S})$ denotes its trace, while $\bm{S}\succeq \bm{0}$ means that $\bm{S}$ is positive semidefinite. For an arbitrary-size matrix $\bm{M}$, ${\text{rank}}(\bm{M})$, $\bm{M}^H$, and $\bm{M}^T$ denote its rank, conjugate transpose, and transpose, respectively. $\bm{I}$ and $\bm{0}$ denote an identity matrix and an all-zero matrix, respectively, with appropriate dimensions. The distribution of a circularly symmetric complex Gaussian (CSCG) random vector with mean vector $\bm{x}$ and covariance matrix $\bm{\Sigma}$ is denoted by $\mathcal{CN}(\bm{x,\Sigma})$; and $\sim$ stands for ``distributed as''. $\mathbb{C}^{x\times y}$ denotes the space of $x\times y$ complex matrices. $\mathbb{R}$ denotes the set of real numbers. ${\mathbb{E}}(\cdot)$ denotes the statistical expectation. $\|\bm{x}\|$ denotes the Euclidean norm of a complex vector $\bm{x}$. $|z|$ and $z^*$ denote the magnitude of a complex number $z$ and its complex conjugate, respectively.

\section{System Model}\label{sec:system1}


\begin{figure}[t]
	\centering
	\setlength{\abovecaptionskip}{2mm}
	\setlength{\belowcaptionskip}{-2mm}
	\includegraphics[width=3in]{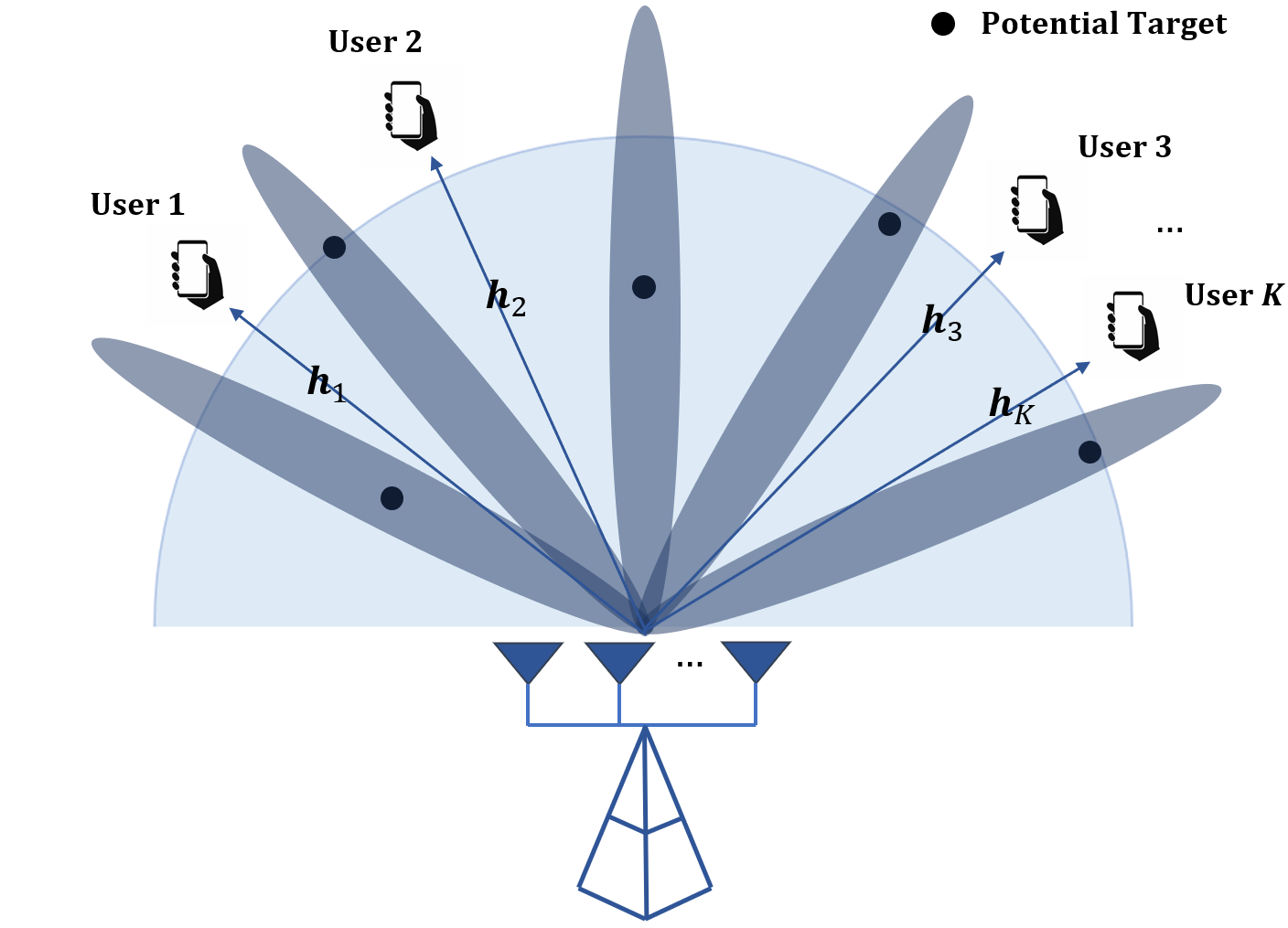}
	\caption{Illustration of the considered downlink ISAC system. } \label{fig:system_model}
\end{figure}


We consider a downlink ISAC system as shown in Fig. \ref{fig:system_model}, in which a BS sends wireless signals to perform radar sensing towards potential targets and downlink communication with $K$ users simultaneously. The communication users then decode their own messages based on the received signals. Meanwhile, as the wireless signals are completely known at the BS, the BS is able to process the echo waves reflected by potential targets for target detection or estimation.

The BS is equipped with a ULA with $N>1$ antennas, and each user is equipped with a single antenna.
In order to perform sensing and communication at the same time with the aid of multiple antennas, we suppose that the BS uses transmit beamforming to send information-bearing signals $s_k$'s for the $K$ users, together with a dedicated radar signal $\mv{s}_0 \in \mathbb{C}^{N \times 1}$. Let $\mathcal{K} \triangleq \{1,...,K\}$ denote the set of communication users. The transmitted information signals $\{s_k\}_{k=1}^K$ are assumed to be independent random variables, with zero mean and unit variance, while the radar signal $\mv{s}_0$ has zero mean and covariance matrix $\mv{R}_d = \mathbb{E}[\mv{s}_0\mv{s}_0^H] \succeq \mv{0}$ and is independent from $\{s_k\}_{k=1}^K$.
Let $\mv{t}_k \in \mathbb{C}^{N \times 1}$ denote the transmit beamforming vector for user $k \in \mathcal{K}$. By combining $s_k$'s and $\mv{s}_0$, the transmit signal $\mv{x} \in \mathbb{C}^{N \times 1}$ by the BS is expressed as
\begin{equation} \label{equ:signal}
	\bm{x} = \sum_{k=1}^K \bm{t}_k s_k + \bm{s}_0.
\end{equation}
Notice that different from the single-beam transmission for information-bearing signals, we consider the general multi-beam transmission for the dedicated radar signal $\mv{s}_0$. In particular, $\mv{s}_0$ can be expressed as
\begin{align}
	\mv{s}_0 = \sum_{i=1}^{N} \mv{w}^{\text r}_i s^{\text r}_i,
\end{align}
where $s^{\text r}_i$ denotes the $i$-th radar waveform that is independent pseudorandom sequences with zero mean and unit variance \cite{sharma2012four}, and $\mv{w}^{\text r}_i$ denotes the corresponding transmit beamforming vector. Accordingly, 
the radar beamformers $\{\mv{w}^{\text r}_i\}_{i=1}^N$ can be determined based on $\bm{R}_d$ via its eigenvalue decomposition (EVD).
Based on (\ref{equ:signal}), the sum transmit power at the BS is given as $\mathbb{E}\left[\| \sum_{k=1}^{K} \mv{t}_k s_k + \mv{s}_0 \|^2\right]   = \sum_{k=1}^{K} \|\mv{t}_k\|^2 + \text{tr}(\mv{R}_d)$. Suppose that the maximum sum transmit power budget at the BS is $P_0$, then we have the following power constraint 
\begin{align}\label{equ:sum_power_constr}
	\sum_{k=1}^{K} \|\mv{t}_k\|^2 + \text{tr}(\mv{R}_d) \leq P_0.
\end{align}

First, we consider the radar target sensing, where the transmit beampattern\footnote{We focus on the transmit beampattern design, as the proper design can result in enhanced sensing performance (in terms of detection, sensing, or recognization) via proper echo wave processing \cite{stoica2007probing}.} is the key performance metric that has been widely adopted in the literature for MIMO radar signal design \cite{stoica2007probing}. The transmit beampattern generally depicts the transmit signal power distribution with respect to the sensing angle $\theta$ ranging from $[-\frac{\pi}{2}, \frac{\pi}{2}]$. In our considered ISAC system, both radar and information signals are jointly used to sense the targets. As a result, the resultant transmit beampattern gain is expressed as
\begin{align}
	\nonumber
	\mathcal{P}(\theta) & = \mathbb{E}\left[ |\mv{a}^H(\theta)(\sum_{k=1}^{K} \mv{t}_k s_k + \mv{s}_0)|^2 \right] \\
	&= \mv{a}^H(\theta) \left(\sum_{k=1}^{K} \mv{t}_k \mv{t}^H_k + \mv{R}_d \right)\mv{a}(\theta),
\end{align}
where
\begin{align} \label{equ:steering}
	\mv{a}(\theta) = [1,e^{j 2\pi \frac{d}{\lambda} \sin \theta},...,e^{j 2\pi \frac{d}{\lambda} (N-1) \sin\theta}]^T
\end{align}
denotes the steering vector at angle $\theta$, with $\lambda$ and $d$ denoting the carrier wavelength and the spacing between two adjacent antennas, respectively. 
In practice, the transmit beampattern is designed based on the radar target sensing requirements. For instance, if we need to perform the detection task without knowing the direction of potential targets, a uniformly distributed beampattern is desired. By contrast, if we roughly know the directions of the targets, e.g., for target tracking, we only need to maximize the beampattern gains towards these interested potential directions \cite{stoica2007probing}.

Next, we consider the information reception at communication users, where their SINRs are used as the performance metric, as widely adopted in wireless communications \cite{heath2018foundations}. Suppose that the channel vector from the BS to each user $i \in \mathcal{K}$ is denoted by $\mv{h}_i$, the received signal at user $i$ is 
\begin{equation} \label{equ:Rx_signal}
	y_i = \mv{h}_i^H \mv{x} + n_i = \mv{h}_i^H (\sum_{k=1}^{K} \mv{t}_k s_k + \mv{s}_0) + n_i,
\end{equation}
where $n_i \sim \mathcal{CN}(0,\sigma_i^2)$ denotes the additive white Gaussian noise (AWGN) at the receiver of user $i$. It is observed in (\ref{equ:Rx_signal}) that each user $i$ suffers from the interference induced by both the information signals $\{s_k\}_{k \neq i}$, as well as the dedicated radar signal $\mv{s}_0$. Notice that $\mv{s}_0$ is pre-determined sequences that can be {\it a-priori} known by both the BS and the users.
Therefore, we consider the following two different types of  communication users, namely Type-I and Type-II receivers, which do not have and do have the capability of cancelling the resultant interference by the radar signals $\mv{s}_0$, respectively\footnote{In order to implement the radar interference cancellation at Type-II receivers, the BS needs to send the information of dedicated radar signals $\bm{s}_0$ to them prior to the ISAC transmission. During the communication, each Type-II receiver $i$ can first estimate the wireless channel $\bm{h}_i$, then cancel $\bm{h}_i^H \bm{s}_0$ from $y_i$, and finally decodes $s_i$ by using the processed signal $y_i - \bm{h}_i^H \bm{s}_0$. }.
\begin{itemize}
	\item \emph{Type-I receivers} (e.g., legacy users) do not have the capability of cancelling the interference of the radar signals before decoding its desirable information signal. In this case, the SINR at receiver $i \in \mathcal{K}$ is
	\begin{align}\label{equ:SINR_Type_I}
		\gamma_{i}^{\text{(I)}} (\{\mv{t}_i\}, \mv{R}_d) =\frac{\left|\mv{h}_i^{H} \mv{t}_i\right|^{2}}{\sum\limits_{k \in \mathcal{K} \atop k \neq i}\left|\mv{h}_i^{H} \mv{t}_{k}\right|^{2} + \mv{h}_i^H \mv{R}_d \mv{h}_i +\sigma_i^2}, \forall i \in \mathcal{K}.
	\end{align}
	\item \emph{Type-II receivers} are dedicatedly designed for
	the ISAC system to have the capability of perfectly cancelling the interference caused by the radar signal $\mv{s}_0$. In this case, the SINR at receiver $i \in \mathcal{K}$ is
	\begin{align}\label{equ:SINR_Type_II}
		\gamma_{i}^{\text{(II)}} (\{\mv{t}_i\})=\frac{\left|\mv{h}_i^{H} \mv{t}_i\right|^{2}}{\sum\limits_{k \in \mathcal{K} \atop k \neq i}\left|\mv{h}_i^{H} \mv{t}_{k}\right|^{2} +\sigma_i^2}, \forall i \in \mathcal{K}.
	\end{align}
\end{itemize}


For the purpose of illustration, Fig. 2(a) and 2(b) show the ISAC system designs with Type-I and Type-II receivers, respectively. For comparison, Fig. 2(c) shows the conventional design without $\mv{s}_0$ \cite{liu2018mu}. It is worth discussing the reasoning behind these three different system setup. As mentioned in \cite{Eldar2020joint}, the setup in Fig. 2(c) cannot leverage the full DoF, especially when the number of users is less than that of antennas, resulting in possible radar beampattern degeneration. From the sensing perspective, the addition of dedicated radar signal will unleash all the available DoF, particularly when the targets are not aligned with the users and dedicated radar beams are needed to illuminate the potential targets. However, from the communication perspective, the dedicated radar signal will, more or less, cause additional interference to the decoding of the communication signals, as can be seen from (\ref{equ:SINR_Type_I}). As the dedicated radar signal can be \textit{a-priori} known by both the BS and the users, it is possible that the users can pre-cancel the radar interference before decoding their own messages. By having such capability at the communication receivers (Type-II receivers), the benefits of adding dedicated radar signals can be fully obtained. This motivates us to consider the ISAC transmit beamforming design with the addition of the dedicated radar signal under Type-II receivers in Fig. 2(b), which has not been considered in the previous literatures, and compare it with the designs under the setup in Fig. 2(a) and Fig. 2(c).

\begin{figure}[t]
	\centering
	\setlength{\abovecaptionskip}{2mm}
	\setlength{\belowcaptionskip}{-4mm}
	\includegraphics[width=3.45in]{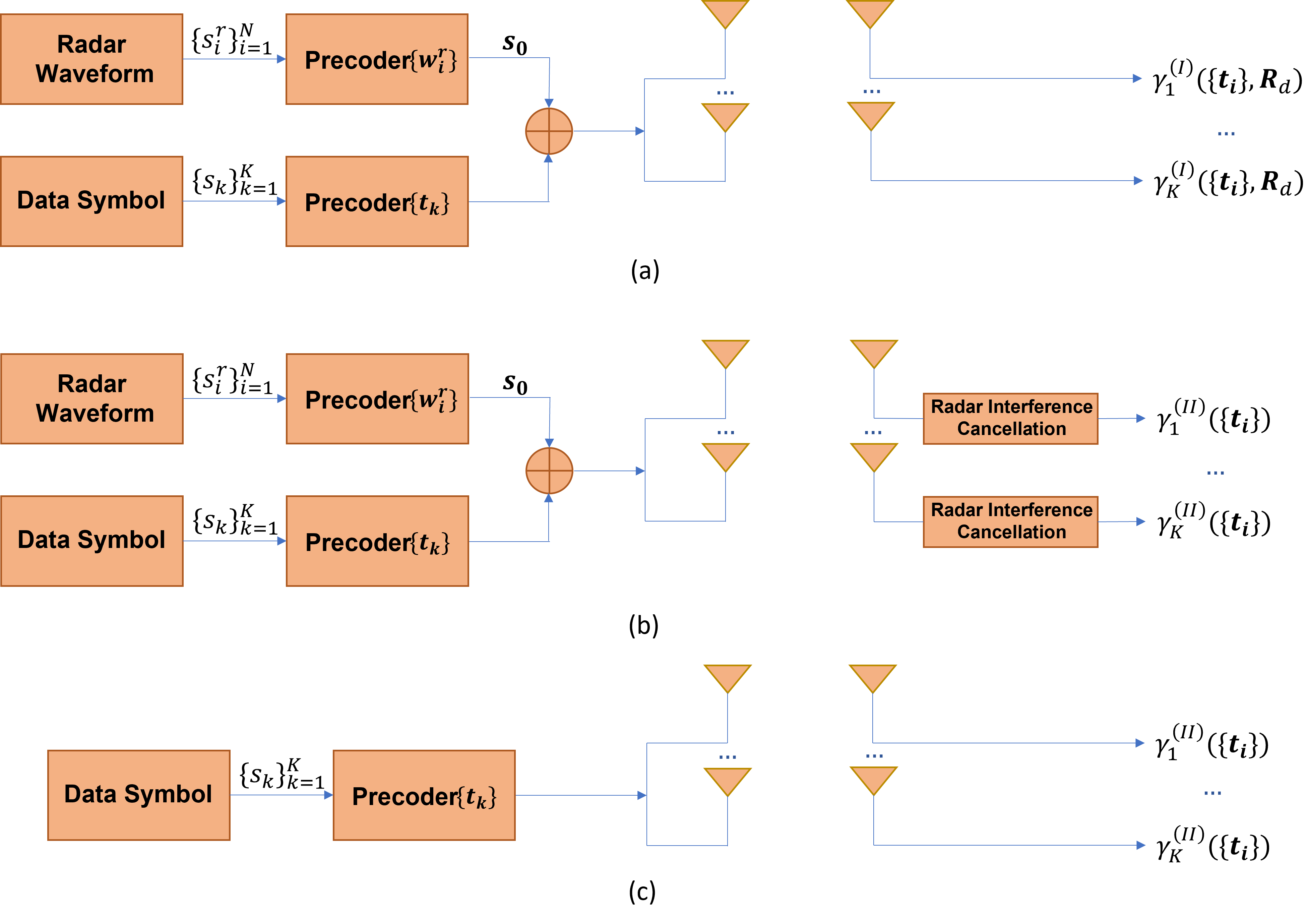}
	\caption{Illustration of different ISAC transceivers designs: (a) Type-I receivers with dedicated radar signals at the BS \cite{Eldar2020joint}; (b) Type-II receivers with dedicated radar signals at the BS; (c) the conventional design without dedicated radar signal at the BS \cite{liu2018mu}.} \label{fig:system_setup}
\end{figure}

It is assumed that the BS perfectly knows all the channel vectors $\mv{h}_i$'s and the desirable beampattern or interested sensing
directions, while each user perfectly knows its own channel vector\footnote{In practice, each user can obtain its channel state information (CSI) via channel estimation, and the BS can obtain the CSI via the users feeding back its estimated CSI in frequency division duplex (FDD) systems, or via itself estimating the reverse link from the users to the BS based on the channel reciprocity in time division duplex (TDD) systems. Notice that there may exist channel estimation and channel quantization errors in the acquired CSI. In this case, robust transmit beamforming designs (see, e.g., \cite{jeong2014beamforming}) can be applied.}.
This assumption is made to characterize the performance upper bounds and gain design insights for practical scenarios when such information is not perfectly known. Under this setup, our objective is to design the joint sensing and communication beamforming (or precoder) to maximize the sensing performance while ensuring the communication requirements at each user, as will be studied in the next two sections in more detail.

\section{SINR-Constrained Beampattern Matching}\label{Section-joint}

In this section, we jointly optimize the information beamforming vectors $\mv{t}_k, \forall k \in \mathcal{K}$,
and the dedicated radar signal covariance matrix $\mv{R}_d$, to minimize the matching error between the overall transmit beampattern and a given desired beampattern, subject to the sum power constraint at the BS and a set of minimum SINR constraints at communication users.

\subsection{Problem Formulation}

Let $\{\tilde{\mathcal{P}}\left(\theta_{m}\right)\}_{m=1}^M$ denote a pre-designed beampattern, which specifies the desired transmit power distribution at the $M$ angles $\{\theta_m\}_{m=1}^M$ in the space. For target detection tasks, $\{\tilde{\mathcal{P}}\left(\theta_{m}\right)\}_{m=1}^M$ can be uniformly distributed \cite{stoica2007probing}, while for target tracking, it can be non-zero \textit{constant} at interested angles with potential targets and zero elsewhere \cite{liu2018mu}. In this case, the beampattern matching error is defined as 
\begin{align}
	\nonumber
	&f_0(\{\mv{t}_k\}, \bm{R}_d, \alpha) \\
	 & = \sum\limits_{m=1}^{M}\left|\alpha \tilde{\mathcal{P}}\left(\theta_{m}\right) - \mv{a}^H(\theta_m) (\sum\limits_{k=1}^{K} \mv{t}_k \mv{t}^H_k + \mv{R}_d) \mv{a}(\theta_m) \right|^{2}.
\end{align}
Here, $\mv{a}\left(\theta_{m}\right)$ denotes the steering vector at direction $\theta_{m}$ as in (\ref{equ:steering}), and $\alpha$ a scaling coefficient, which is introduced to adjust the scaling level of $\tilde{\mathcal{P}}(\theta_m)$ such that the transmit beampattern can better match the scaled version of the desired beampattern $\tilde{\mathcal{P}}(\theta_m)$, as commonly adopted in MIMO radar design \cite{li2008mimo}.

Our objective is to minimize the beampattern matching error, for which the problem is formulated as follows with Type-I and Type-II receivers, respectively.
\begin{subequations}
	\begin{align}
		(\text{P}1): &\min\limits_{\{\bm{t}_k\}, \bm{R}_d, \alpha} f_0(\{\mv{t}_k\}, \bm{R}_d, \alpha) \\
		\text { s.t. } & \gamma_{i}^{\text{(I)}}(\{\mv{t}_i\}, \mv{R}_d) \geq \Gamma_{i}, \quad \forall i \in \mathcal{K}\\
		\label{equ:SINR-I}
		&\quad \quad \sum_{k=1}^{K} ||\mv{t}_k||^2  + \text{tr}(\mv{R}_d) = P_0\\
		\label{equ:sum-P1}
		&\quad \quad \mv{R}_d \succeq 0.
	\end{align}
\end{subequations}
and 
\begin{subequations}
	\begin{align}
		(\text{P}2): &\min\limits_{\{\bm{t}_k\}, \bm{R}_d, \alpha} f_0(\{\mv{t}_k\}, \bm{R}_d, \alpha) \\
		\text { s.t. } & \gamma_{i}^{\text{(II)}}(\{\mv{t}_i\}) \geq \Gamma_{i}, \quad \forall i \in \mathcal{K}\\ 
		\nonumber
		&\quad \quad (\text{\ref{equ:SINR-I}}) \text{ and } (\text{\ref{equ:sum-P1}}).
	\end{align}
\end{subequations}
Here, $\Gamma_{i} \ge 0$ denote the minimum SINR requirement at each receiver $i \in \mathcal{K}$.

In problems (P1)\footnote{(P1) is different from that in \cite{Eldar2020joint} in the following two aspects. First, we consider the beampattern matching error as the sole objective function so that a fair analysis and comparison could be conducted involving \cite{liu2018mu}, \cite{Eldar2020joint} and our design. Second, this paper considers the sum-power constraint at the BS versus the per-antenna power constraints in \cite{Eldar2020joint} since it has been widely adopted in wireless communication networks.} and (P2), we consider the equality power constraints in (\ref{equ:SINR-I}) in order for the BS to use up all the transmit power to maximize the radar performance for target tracking or detection \cite{li2007mimo}. 



\subsection{Optimal Solution via SDR}

Notice that both problems (P1) and (P2) are non-convex and thus are difficult to be optimally solved in general. Fortunately, we show that one can apply the SDR technique \cite{luo2010semidefinite} to obtain their \textit{optimal} solutions in this subsection. Towards this end, we introduce new auxilary variables  $\mv{T}_k = \mv{t}_k \mv{t}_k^H, \forall k \in \mathcal{K}$, where  $\mv{T}_k \succeq 0$ and $\text{rank}(\mv{T}_k) = 1$. In this case, let $f_1(\{\bm{T}_k\}, \bm{R}_d, \alpha)$ denote
\begin{align} \label{equ:f1}
	 \sum\limits_{m=1}^{M}\left|\alpha \tilde{\mathcal{P}}\left(\theta_{m}\right) - \mv{a}^H(\theta_m) (\sum\limits_{k=1}^{K} \mv{T}_k + \mv{R}_d) \mv{a}(\theta_m) \right|^{2}.
\end{align}
Problems (P1) and (P2) are equivalently reformulated as (P1.1) and (P2.1) in the following, respectively.
\begin{subequations}
	\begin{align}
		(\text{P}1.1): & \min\limits_{\{\bm{T}_k\}, \bm{R}_d, \alpha} f_1(\{\bm{T}_k\}, \bm{R}_d, \alpha)\\
		\nonumber
		\text { s.t. } &\frac{\text{tr}\left(\bm{h}_{i} \bm{h}_{i}^H \bm{T}_{i}\right)}{\Gamma_{i}}-\sum\limits_{k \in \mathcal{K} \atop k \neq i} \text{tr}\left(\bm{h}_{i} \bm{h}_{i}^H \bm{T}_{k}\right)- \\
		& \quad \quad \text{tr}\left(\bm{h}_{i} \bm{h}_{i}^H \bm{R}_{d}\right)-\sigma_{i}^{2} \geq 0, \forall i \in \mathcal{K}\\
		\label{equ:T_power}
		\quad \quad & \sum_{k=1}^{K} \text{tr}(\mv{T}_k)  + \text{tr}(\mv{R}_d) = P_0\\
		\label{equ:rank}
		\quad \quad &\mv{R}_d \succeq 0, \mv{T}_k \succeq 0, \text{rank}(\mv{T}_k) = 1, \forall k \in \mathcal{K}.
	\end{align}
\end{subequations}
\begin{subequations}
	\begin{align}
		(\text{P}2.1): &\min\limits_{\{\bm{T}_k\}, \bm{R}_d, \alpha} f_1(\{\bm{T}_k\}, \bm{R}_d, \alpha) \\
		\text { s.t. } &\frac{\text{tr}\left(\bm{h}_{i} \bm{h}_{i}^H \bm{T}_{i}\right)}{\Gamma_{i}}-\sum\limits_{k \in \mathcal{K} \atop k \neq i} \text{tr}\left(\bm{h}_{i} \bm{h}_{i}^H \bm{T}_{k}\right)
		-\sigma_{i}^{2} \geq 0, \forall i \in \mathcal{K} \\ \nonumber
		\quad \quad & \text{(\ref{equ:T_power})} \text{ and } \text{(\ref{equ:rank})}.
	\end{align}
\end{subequations}

However, problems (P1.1) and (P2.1) are still non-convex due to the rank-one constraints on $\mv{T}_k$'s. To deal with this issue, we relax the rank-one constraints and accordingly get the SDRs of problems (P1.1) and (P2.1), denoted as (SDR1) and (SDR2), respectively. Notice that both (SDR1) and (SDR2) are convex QSDPs and thus can be solved optimally by convex optimization solvers such as CVX \cite{grant2014cvx}.
Notice that if the ranks of the obtained communication covariance matrices $\mv{T}_k$'s in the optimal solution to (SDR1) (or (SDR2)) are all equal to 1, then they are also optimal for (P1.1) (or (P2.1)). Otherwise, we should further perform the Gaussian randomization \cite{luo2010semidefinite} to construct rank-one solutions that are generally sub-optimal. Fortunately, the following propositions show that with the addition of dedicated radar signal, there always exist optimal rank-one solutions to (SDR1) and (SDR2), and thus the Gaussian randomization is not needed.


\begin{proposition}\label{T_1} \emph{
		There always exists a globally optimal solution to problem (SDR1), denoted as $\{\{\tilde{\mv{T}}_k\},\tilde{\mv{R}}_d,\tilde{\alpha}\}$, such that
		\begin{align}
			\nonumber
			\text{rank}(\tilde{\mv{T}}_k) = 1, \forall k \in \mathcal{K}.
	\end{align}}
	\begin{proof}
		Notice that problem (P1.1) has a similar structure as the joint information and radar beamforming problem in \cite{Eldar2020joint}, (30) and can be similarly proved as in \cite[Theorem 1]{Eldar2020joint},
		for which the details are omitted for brevity.
	\end{proof}
\end{proposition}


\begin{proposition}\label{T_2} \emph{
		There always exists a globally optimal solution to problem (SDR2), denoted as $\{\{\bar{\mv{T}}_k\},\bar{\mv{R}}_d,\bar{\alpha}\}$, such that
		\begin{align}
			\nonumber
			\text{rank}(\bar{\mv{T}}_k) = 1, \forall k \in \mathcal{K}.
	\end{align}}
	\begin{proof}
		See Appendix \ref{Proof_theorem P2}.
	\end{proof}
\end{proposition}
\begin{remark} \label{remark:Joint} \emph{
		By comparing problems (P1.1) and (P2.1), it is observed that every feasible solution to problem (P1.1) is also feasible to problem (P2.1) but not vice versa. Therefore, it is evident that problem (P2.1) always yields an equal or smaller beampattern matching error than (P1.1) due to the enlarged feasible region.} 
\end{remark}

\subsection{Analytic Performance Comparison versus Corresponding Design without Radar Signals}\label{Section_nec_first}

To gain more insights, this subsection analyzes the performance achieved by the joint beamforming designs in (P1) (or (P1.1)) and (P2) (or (P2.1)) for Type-I and Type-II receivers, as compared with the corresponding design without radar signals. We also show the necessity of radar signals under different cases. 

To facilitate the comparison, let ${f_{\text{(P1.1)}}^\star}$, ${f_{\text{(SDR1)}}^\star}$, ${f_{\text{(SDR2)}}^\star}$, and ${f_{\text{(P2.1)}}^\star}$ denote the optimal objective function values achieved by solving (P1.1), (SDR1), (SDR2), and (P2.1), respectively. Then we have 
\begin{align} \label{equ:first_half_fund}
	{f_{\text{(P1.1)}}^\star} \stackrel{(a)}{=} {f_{\text{(SDR1)}}^\star} \stackrel{(b)}{\geq} {f_{\text{(SDR2)}}^\star} \stackrel{(c)}{=} {f_{\text{(P2.1)}}^\star},
\end{align}
where (a), (b), and (c) follow from Proposition 1, Remark 1, and Proposition 2, respectively.



Next, we introduce the benchmark beampattern matching problem (P3) without dedicated radar signals, which corresponds to a special case of (P1) and (P2) by setting $\mv{R}_d = \boldsymbol{0}$. In this case, by introducing $\mv{T}_k = \mv{t}_k \mv{t}_k^H$ with $ \mv{T}_k \succeq \mv 0$, $\text{rank}(\mv{T}_k) \le 1$, $ \forall k \in \mathcal{K}$, and let $f_3(\{\bm{T}_k\}, \alpha)$ denote
\begin{align} \label{equ:f3}
	\sum\limits_{m=1}^{M}\left|\alpha \tilde{\mathcal{P}}\left(\theta_{m}\right) - \mv{a}^H(\theta_m) (\sum\limits_{k=1}^{K} \mv{T}_k) \mv{a}(\theta_m) \right|^{2},
\end{align}
 we equivalently re-express problem (P3) as
\begin{align} \label{equ:P3.1}
	(\text{P}3.1): &\min\limits_{\{\bm{T}_k\}, \alpha} f_3(\{\bm{T}_k\}, \alpha) \\
	\nonumber
	\text { s.t. } & \frac{\text{tr}\left(\bm{h}_{i} \bm{h}_{i}^H \bm{T}_{i}\right)}{\Gamma_{i}}-\sum\limits_{k \in \mathcal{K} \atop k \neq i} \text{tr}\left(\bm{h}_{i} \bm{h}_{i}^H \bm{T}_{k}\right)
	-\sigma_{i}^{2} \geq 0, \forall i \in \mathcal{K} \\
	\nonumber
	&  \sum_{k=1}^{K} \text{tr}(\mv{T}_k) = P_0\\
	\nonumber
	&  \mv{T}_k \succeq 0, \text{rank}(\mv{T}_k) = 1, \forall k \in \mathcal{K}.
\end{align}
By removing the rank-one constraints on $\mv{T}_k$'s, we have the SDR of (P3.1) as (SDR3). Let ${f_{\text{(P3.1)}}^\star}$ and ${f_{\text{(SDR3)}}^\star}$ denote the optimal objective function values achieved by (P3.1) and (SDR3), respectively, where we then have the following proposition.

\begin{proposition}\label{T_30} \emph{
		Let $\{\{\tilde{\mv{T}}_k\},\tilde{\mv{R}}_d,\tilde{\alpha}\}$ denote the optimal solution to problem (SDR1). Then we can construct $\{\{\hat{\mv{T}}_k\},\hat{\alpha}\}$ as follows, which are optimal for problem (SDR3).
		\begin{align}
			&\hat{\bm{T}}_{k} = \tilde{\bm{T}}_{k} + \beta_k \tilde{\mv{R}}_d, \forall k \in \mathcal{K},\\
			&\hat{\alpha} = \tilde{\alpha},
		\end{align}
		where $\{\beta_k\}$ are any arbitrary real numbers satisfying that $\sum_{k=1}^K \beta_k = 1$,  and $\beta_k \geq 0, \forall k \in \mathcal{K}$, such that
		\begin{align}
			\sum\limits_{k=1}^K \tilde{\bm{T}}_{k} + \tilde{\mv{R}}_d = \sum\limits_{k=1}^{K} \hat{\bm{T}}_{k}.
		\end{align}
		Accordingly, the optimal value of (SDR3) is same as that of (SDR1), i.e.,
		\begin{align} \label{equ:second_half_fund}
			{f_{\text{(SDR3)}}^\star} = {f_{\text{(SDR1)}}^\star}.
		\end{align}
	}
	\begin{proof}	
		See Appendix \ref{Proof_theorem P3}.
	\end{proof}
\end{proposition}
Combining (\ref{equ:first_half_fund}), (\ref{equ:second_half_fund}), and the fact that ${f_{\text{(P3.1)}}^\star} \geq {f_{\text{(SDR3)}}^\star}$, we have
\begin{align} \label{equ:Fund_1}
	{f_{\text{(P3.1)}}^\star} \geq {f_{\text{(SDR3)}}^\star} = {f_{\text{(SDR1)}}^\star} = {f_{\text{(P1.1)}}^\star} \geq {f_{\text{(P2.1)}}^\star} = {f_{\text{(SDR2)}}^\star}.
\end{align}
It follows from (\ref{equ:Fund_1}) that $f^\star_{\text{(P3.1)}} \geq f^\star_{\text{(P1.1)}} \geq f^\star_{\text{(P2.1)}}$. This shows that with both types of receivers, adding dedicated radar signals generally improve the system performance in terms of lower beampattern matching errors, thanks to the exploitation of full DoF (similarly as that in \cite{Eldar2020joint}). It is also shown that  the case with Type-II receivers outperforms that with Type-I receivers, thanks to the employment of radar interference cancellation.

Besides the above general channel conditions, it is also interesting to discuss a special case with LOS communication channels, for which we have the following proposition.

\begin{proposition}\label{T_LOS} \emph{
		When the wireless channels from the BS to the communication users are LOS (i.e., $\mv{h}_i$ is given in the form of $\mv{h}_i = [1,e^{j \phi_i},...,e^{j (N-1)\phi_i}]^T$ with $\phi_i = 2 \pi \frac{d}{\lambda} \sin(\theta_i)$, where $\lambda$ and $d$ denotes the carrier wavelength and the spacing between two adjacent antennas, respectively, and $\theta_i$ denotes the direction of the receivers), the SDR of problem (P3.1) (or equivalent (P3)) is tight, i.e., ${f_{\text{(P3.1)}}^\star} = {f_{\text{(SDR3)}}^\star}$. }
	\begin{proof}
		See Appendix \ref{Proof_theorem_LOS}.
	\end{proof}
\end{proposition}

By combining Propositions \ref{T_1}, \ref{T_30}, and \ref{T_LOS}, it follows that $f_{\text{(P3.1)}}^\star = f_{\text{(SDR3)}}^\star = f_{\text{(SDR1)}}^\star =  f_{\text{(P1.1)}}^\star$. In other words, problem (P1) with radar signals and type-I receivers achieves the same optimal matching error as that by problem (P3) without dedicated radar signals. This shows that with Type-I receivers, the radar signals are not needed in this special case with LOS communication channels, i.e., the DoFs provided by information signals are sufficient for ISAC in this case.

\section{SINR-Constrained Minimum Weighted Beampattern Gain Maximization}\label{S-maxmin}

In this section, we propose an alternative radar sensing design criterion termed minimum weighted  beampattern gain maximization, based on which our objective is to maximize the minimum weighted  beampattern gain illuminated in desirable directions of potential sensing targets, subject to the SINR constraints at individual communication users. We will show that this design leads to a better sensing beampattern at lower computation complexity than the conventional beampattern matching design in Section \ref{Section-joint}.

\subsection{Problem Formulation}

In this design, our objective is to maximize the minimum weighted beampattern gain at a given set of interested angles $\Theta \triangleq \{\theta_1,\theta_2,...,\theta_Q\}$, where $Q$ denotes the number of quantized angles of interest. For instance, for target detection tasks without knowing any prior information about targets' potential locations, $\{\theta_q\}_{q=1}^Q$ may correspond to uniformly quantized angles ranging from $-\frac{\pi}{2}$ to $\frac{\pi}{2}$. For target tracking tasks, the interested angles can be specified based on the targets' potential locations (similarly as the angles with non-zero entries in the pre-designed beampattern $\tilde{\mathcal{P}}(\theta_m)$ in Section \ref{Section-joint}). In this case, the SINR-constrained minimum weighted beampattern gain maximization problems are formulated as (P4) and (P5) in the following, by considering Type-I and Type-II communication receivers, respectively.
\begin{subequations}
	\begin{align}
		(\text{P}4): &\max \limits_{t,\bm{R}_d,\{\bm{t}_k\}}  t \\
		\text { s.t. } & \mv{a}^{H}\left(\theta\right) \left(\mv{R}_d + \sum\limits_{k=1}^{K} \mv{t}_k \mv{t}_k^H \right) \mv{a} \left(\theta\right) \geq \eta_\theta t, \forall \theta \in \Theta\\
		&\quad \quad \gamma_{i}^{\text{(I)}}(\{\mv{t}_i\}, \mv{R}_d) \geq \Gamma_{i}, \quad \forall i \in \mathcal{K}\\
		\label{equ:sum-P5}
		&\quad \quad \sum_{k=1}^{K} ||\mv{t}_k||^2  + \text{tr}(\mv{R}_d) \leq P_0\\
		\label{equ:psd-P5}
		&\quad \quad \mv{R}_d \succeq 0.
	\end{align}
\end{subequations}
\begin{subequations}
	\begin{align}
		(\text{P}5): &\max \limits_{t,\bm{R}_d,\{\bm{t}_k\}}  t \\
		\text { s.t. } & \mv{a}^{H}\left(\theta\right) \left(\mv{R}_d + \sum\limits_{k=1}^{K} \mv{t}_k \mv{t}_k^H \right) \mv{a} \left(\theta\right) \geq \eta_\theta t, \forall \theta \in \Theta \\
		&\quad \quad \gamma_{i}^{\text{(II)}}(\{\mv{t}_i\}) \geq \Gamma_{i}, \quad \forall i \in \mathcal{K}\\
		\nonumber
		&\quad \quad \text{(\ref{equ:sum-P5})} \text{ and } \text{(\ref{equ:psd-P5})}.
	\end{align}
\end{subequations}
where $t$ denotes the minimum weighted beampattern gain and $\eta_\theta$ denotes the beampattern gain weight at each interested sensing angle in $\Theta$, which characterizes the transmit power density over space to meet certain sensing requirements.
Notice that problems (P4) and (P5) are both non-convex. In the following, we will adopt the SDR technique to obtain their \textit{optimal} solutions. 

\subsection{Optimal Solution via SDR}

Similarly as in Section \ref{Section-joint}, we define $\mv{T}_k = \mv{t}_k \mv{t}_k^H$ with $ \mv{T}_k \succeq \mv 0$ and $\text{rank}(\mv{T}_k) \le 1$, $ \forall k \in \mathcal{K}$. Accordingly, problems (P4) and (P5) can be reformulated as (P4.1) and (P5.1) as follows, respectively.
\begin{subequations}
	\begin{align}
		(\text{P}4.1): &\max \limits_{t,\bm{R}_d,\{\bm{T}_k\}} t  \\ 
		\text { s.t. } & \mv{a}^{H}\left(\theta\right) \left(\mv{R}_d + \sum\limits_{k=1}^{K} \mv{T}_k\right) \mv{a} \left(\theta\right) \geq \eta_\theta t, \forall \theta \in \Theta \\
		\nonumber
		& \frac{\text{tr}\left(\bm{h}_{i} \bm{h}_{i}^H \bm{T}_{i}\right)}{\Gamma_{i}}-\sum\limits_{k \in \mathcal{K} \atop k \neq i} \text{tr}\left(\bm{h}_{i} \bm{h}_{i}^H \bm{T}_{k}\right) \\
		& \quad \quad -\text{tr}\left(\bm{h}_{i} \bm{h}_{i}^H \bm{R}_{d}\right)-\sigma_{i}^{2} \geq 0, \forall i \in \mathcal{K} \\
		\label{equ:sum-P41}
		&	\text{tr}\left(\sum\limits_{k=1}^{K} \mv{T}_k + \mv{R}_d\right) \leq P_{0} \\
		\label{equ:psd-P41}
		& \mv{R}_d \succeq 0, \mv{T}_k \succeq 0, \text {rank}\left(\mv{T}_k\right) = 1, \quad \forall k \in \mathcal{K}.
	\end{align}
\end{subequations}
\begin{subequations}
	\begin{align}
		(\text{P}5.1): &\max \limits_{t,\bm{R}_d,\{\bm{T}_k\}} t  \\ 
		\text { s.t. } & \mv{a}^{H}\left(\theta\right) \left(\mv{R}_d + \sum\limits_{k=1}^{K} \mv{T}_k\right) \mv{a} \left(\theta\right) \geq \eta_\theta t, \forall \theta \in \Theta\\ 
		& \frac{\text{tr}\left(\bm{h}_{i} \bm{h}_{i}^H \bm{T}_{i}\right)}{\Gamma_{i}}-\sum\limits_{k \in \mathcal{K} \atop k \neq i} \text{tr}\left(\bm{h}_{i} \bm{h}_{i}^H \bm{T}_{k}\right)
		-\sigma_{i}^{2} \geq 0, \forall i \in \mathcal{K}\\
		\nonumber
		& \text{(\ref{equ:sum-P41})} \text{ and } \text{(\ref{equ:psd-P41})}.
	\end{align}
\end{subequations}
\begin{remark} \label{remark_3} \emph{
		Similar as in Remark \ref{remark:Joint}, it is evident that (P5)/(P5.1) with Type-II receivers always yields an equal or  greater minimum weighted beampattern gains at interested angles than (P4)/(P4.1) with Type-I receivers, due to the enlarged feasible region.}
\end{remark}

However, problems (P4.1) and (P5.1) are still non-convex due to the rank-one constraints on $\bm{T}_k$'s. To resolve this issue, we relax the rank-one constraints and obtain the SDRs of (P4.1) and (P5.1) as (SDR4) and (SDR5), respectively, which are both separable semidefinite programs (SSDPs) that can be solved optimally by CVX. We proceed to show that the two SDRs are indeed tight in the following proposition, based on which the global optimal solutions to (P4.1) and (P5.1) can be obtained.
\begin{proposition}\label{T_56} \emph{
		There always exist globally optimal solutions to problems (SDR4) and (SDR5), which are denoted as $\{\{\tilde{\mv{T}}_k\},\tilde{\mv{R}}_d\}$ and $\{\{\bar{\mv{T}}_k\},\bar{\mv{R}}_d\}$, respectively, such that
		\begin{align}
			\nonumber
			\text{rank}(\tilde{\mv{T}}_k) = 1, \forall k \in \mathcal{K}, \quad \text{rank}(\bar{\mv{T}}_k) = 1, \forall k \in \mathcal{K}.
	\end{align}}
	\begin{proof}
		Similar as the proof for Propositions \ref{T_1} and \ref{T_2}, suppose that $\{\{{\mv{T}}_k^\star\},\mv{R}_d^\star\}$ (or $\{\{\hat{\mv{T}}_k\},\hat{\mv{R}}_d\}$)  is one optimal solution to (SDR4) (or (SDR5)), which does not necessarily meet the rank-one constraints. As shown in Appendix \ref{Proof_theorem P2}, we can always construct an alternative optimal solution $\{\{\tilde{\mv{T}}_k\},\tilde{\mv{R}}_d\}$ (or $\{\{\bar{\mv{T}}_k\},\bar{\mv{R}}_d     \}$)  meeting the rank-one constraints, such that the same minimum weighted beampattern gains can be achieved by $\{\{\tilde{\mv{T}}_k\},\tilde{\mv{R}}_d\}$ (or $\{\{\bar{\mv{T}}_k\},\bar{\mv{R}}_d\}$). As the proof procedure is similar as that in Appendix \ref{Proof_theorem P2}, the details are skipped for brevity.
	\end{proof}
\end{proposition}

\subsection{Analytic Performance Comparison versus Corresponding Design without Radar Signals} \label{Section_Nec_second}

To gain more insights, this subsection analyzes the performance achieved by the newly proposed minimum weighted beampattern gain maximization beamforming designs in (P4) (or (P4.1)) and (P5) (or (P5.1)) for Type-I and Type-II receivers, as compared with the corresponding design without radar signals. We also show the necessity of radar signals under different cases.

To facilitate the comparison, let $t_{\text{(SDR5)}}^\star$, $t_{\text{(P5.1)}}^\star$, $t_{\text{(P4.1)}}^\star$, and $t_{\text{(SDR4)}}^\star$ denote the optimal objective function values achieved by solving (SDR5), (P5.1), (P4.1), and (SDR4), respectively. Then we have
\begin{align}\label{equ:first_part_fund2}
	t_{\text{(SDR5)}}^\star \stackrel{(e)}{=} t_{\text{(P5.1)}}^\star \stackrel{(f)}{\geq} t_{\text{(P4.1)}}^\star \stackrel{(g)}{=} t_{\text{(SDR4)}}^\star
\end{align}
where (e) and (g) follow from Proposition 5 and (f) holds due to Remark 2. 

Next, we introduce the corresponding benchmark design problem (P6) without dedicated radar signals, which corresponds to problem (P4)/(P5) by setting $\mv{R}_d= \boldsymbol{0}$. In this case, by introducing $\mv{T}_k = \mv{t}_k \mv{t}_k^H$ with $ \mv{T}_k \succeq \mv 0$ and $\text{rank}(\mv{T}_k) \le 1$, $ \forall k \in \mathcal{K}$, we re-express (P6) as
\begin{align}\label{{equ:P6.1}}
	(\text{P}6.1): &\max \limits_{t,\{\bm{T}_k\}} t  \\ \nonumber
	\text { s.t. } & \mv{a}^{H}\left(\theta\right) \left(\sum\limits_{k=1}^{K} \mv{T}_k\right) \mv{a} \left(\theta\right) \geq \eta_\theta t, \forall \theta \in \Theta \\ \nonumber
	&\frac{\text{tr}\left(\bm{h}_{i} \bm{h}_{i}^H \bm{T}_{i}\right)}{\Gamma_{i}}-\sum\limits_{k \in \mathcal{K} \atop k \neq i} \text{tr}\left(\bm{h}_{i} \bm{h}_{i}^H \bm{T}_{k}\right)
	-\sigma_{i}^{2} \geq 0, \forall i \in \mathcal{K}\\\nonumber
	&\text{tr}\left(\sum\limits_{k=1}^{K} \mv{T}_k\right) \leq P_{0} \\\nonumber
	&\mv{T}_k \succeq 0, \text {rank}\left(\mv{T}_k\right) = 1, \quad \forall k \in \mathcal{K}.
\end{align}

By removing the rank-one constraints on $\mv{T}_k$'s, we have the SDR of (P6.1) as (SDR6). Let $t_{\text{(P6.1)}}^\star$ and $t_{\text{(SDR6)}}^\star$ denote the optimal objective function values achieved by solving (P6.1) and (SDR6), respectively, where we have the following proposition that can be similarly proved as in Proposition \ref{T_30}.
\begin{proposition}\label{T_60} \emph{
		Let $\{\{\tilde{\mv{T}}_k\},\tilde{\mv{R}}_d\}$ denote the optimal solution to problem (SDR4). Then we can construct $\{\{\hat{\mv{T}}_k\}\}$ as follows, which are optimal for problem (SDR6).
		\begin{align}
			\hat{\bm{T}}_{k} = \tilde{\bm{T}}_{k} + \beta_k \tilde{\mv{R}}_d, \forall k \in \mathcal{K},
		\end{align}
		where $\{\beta_k\}$ are any arbitrary real numbers satisfying that $\sum_{k=1}^K \beta_k = 1,$, and $\beta_k \geq 0, \forall k \in \mathcal{K},$ such that
		\begin{align}
			\sum\limits_{k=1}^K \tilde{\bm{T}}_{k} + \tilde{\mv{R}}_d = \sum\limits_{k=1}^{K} \hat{\bm{T}}_{k}.
		\end{align}
		Accordingly, the optimal value of (SDR6) is same as that of (SDR4), i.e.,
		\begin{align} \label{equ:second_part_fund2}
			t_{\text{(SDR6)}}^\star = t_{\text{(SDR4)}}^\star.
	\end{align}}
\end{proposition}
Combining (\ref{equ:first_part_fund2}), (\ref{equ:second_part_fund2}), and the fact that $t_{\text{(SDR6)}}^\star \geq t_{\text{(P6.1)}}^\star$, we have
\begin{align} \label{equ:Fund_2}
	t_{\text{(SDR5)}}^\star = t_{\text{(P5.1)}}^\star \geq t_{\text{(P4.1)}}^\star = t_{\text{(SDR4)}}^\star = t_{\text{(SDR6)}}^\star \geq t_{\text{(P6.1)}}^\star.
\end{align}
It thus follows from (\ref{equ:Fund_2}) that $t_{\text{(P5.1)}}^\star \geq t_{\text{(P4.1)}}^\star \geq t_{\text{(P6.1)}}^\star$. Notice that the result in (\ref{equ:Fund_2}) is similar to that in (\ref{equ:Fund_1}) for a different design criterion of beampattern matching. This shows that with both types of receivers, adding dedicated radar signals generally improves the system performance in terms of higher minimum weighted beampattern gain, thanks to the exploitation of full DoF. It is also shown that the case with Type-II receivers outperforms that with Type-I receivers, thanks to the employment of radar interference cancellation.

Besides the above general channel conditions, it is also interesting to discuss a special case with LOS communication channels, for which we have the following proposition, which can also be similarly proved as in Proposition \ref{T_LOS}.

\begin{proposition}\label{T_LOS2} \emph{
		When the wireless channels from the BS to the communication users are LOS (i.e., $\mv{h}_i$ is given in the form of $\mv{h}_i = [1,e^{j \phi_i},...,e^{j (N-1)\phi_i}]^T$ with $\phi_i = 2 \pi \frac{d}{\lambda} \sin(\theta_i)$, where $\theta_i$ denotes the direction of the receivers), the SDR of problem (P6.1) (or equivalent (P6)) is tight, i.e., $t^\star_{\text{(P6.1)}} = t^\star_{\text{(SDR6)}}$. }
\end{proposition}

By combining Propositions \ref{T_56}, \ref{T_60} and \ref{T_LOS2}, it follows that $t_{\text{(P6.1)}}^\star = t_{\text{(SDR6)}}^\star = t_{\text{(SDR4)}}^\star = t_{\text{(P4.1)}}^\star$. In other words, problem (P4) with dedicated radar signals and type-I receivers achieves the same minimum weighted beampattern gain as that by problem (P6) without dedicated radar signals. This shows that with Type-I receivers, the dedicated radar signals are not needed in this special case with LOS communication channels, i.e., the DoFs provided by information signals are sufficient for ISAC in this case.

\subsection{Complexity Analysis} \label{Section_complexity}

In this subsection, we analyze the computational complexity for solving problem (SDR1) /(SDR2) versus that for (SDR4)/(SDR5) in order to show the benefit of the newly proposed minimum weighted beampattern gain maximization design. Notice that problems (SDR1) and (SDR2) are both QSDPs. Given a solution accuracy $\epsilon$, the worst-case complexity to solve the QSDP problems with the primal-dual interior-point method is $\mathcal{O}(K^{6.5}N^{6.5}\log(1/\epsilon))$ \cite{Eldar2020joint}. By contrast, problems (SDR4) and (SDR5) are SSDPs. It follows from \cite{luo2010semidefinite} that given a solution accuracy $\epsilon$, the worst-case complexity to solve the SSDP with the primal-dual interior-point method is $\mathcal{O}( \max \{KN+N+1,Q+K+1\}^4 (KN+N+1)^{1/2} \log(1/\epsilon)) \approx \mathcal{O}( K^{4.5} N^{4.5} \log(1/\epsilon)) $, which is significantly less than that for solving the QSDPs (SDR1) and (SDR2). This shows the benefit of the proposed minimum weighted beampattern gain maximization design over its beampattern matching counterpart, in terms of the computational complexity.

\section{Numerical Results} \label{Section_results}

In this section, we provide numerical results to validate the performance of our proposed transmit beamforming designs under beampattern matching and minimum weighted beampattern gain maximization criteria together with both Type-I and Type-II receivers. 

\begin{figure}[t]
	\centering
	\setlength{\abovecaptionskip}{-4mm}
	\setlength{\belowcaptionskip}{-4mm}
	\setlength{\abovecaptionskip}{+3pt}
	\subfigure[The case with Rayleigh fading channels]{ \label{fig:A-I-Rich-a}
		\includegraphics[width=2.6in]{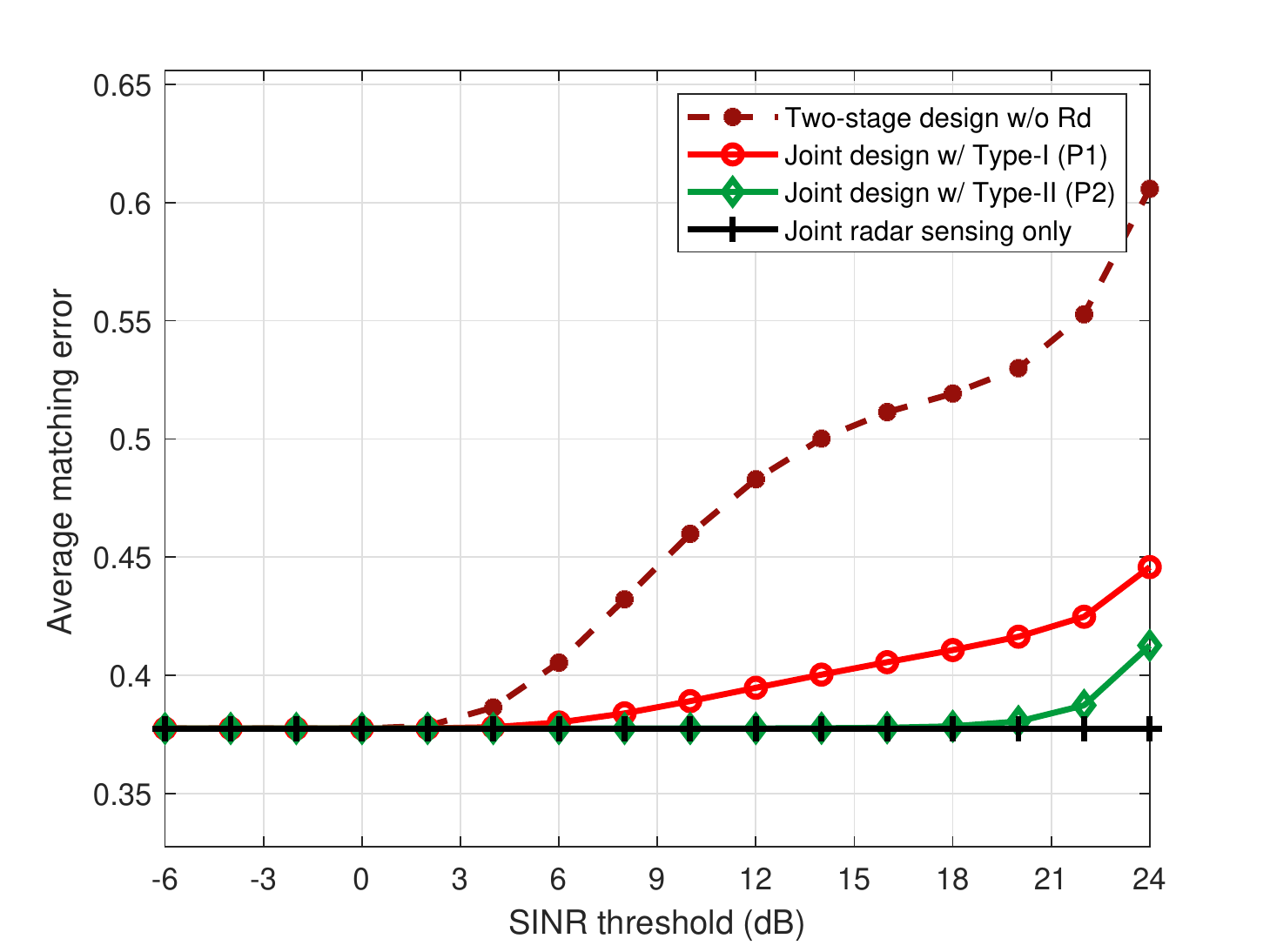}}
	\subfigure[The case with LOS channels]{ \label{fig:A-I-Rich-b}
		\includegraphics[width=2.6in]{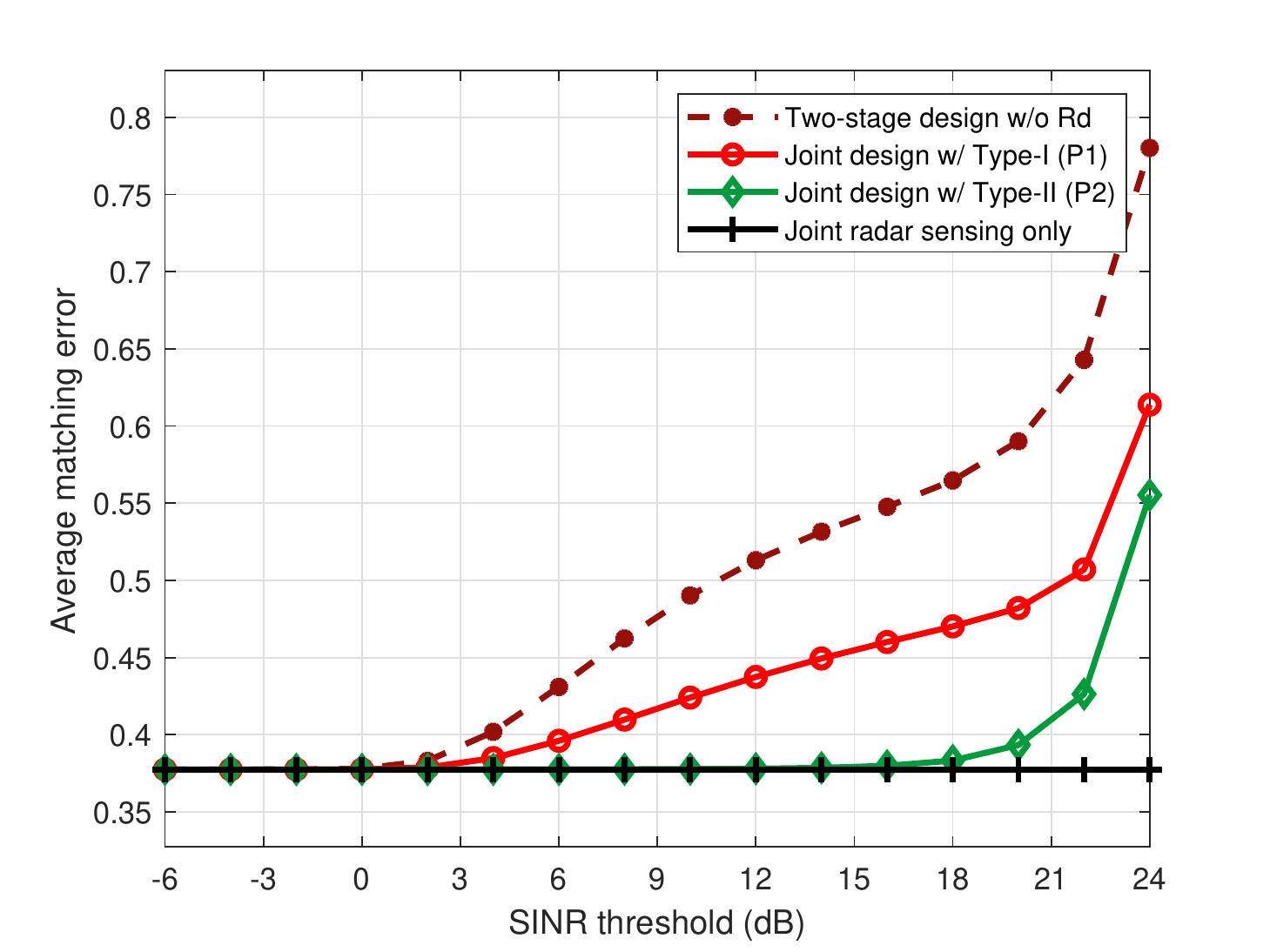}}
	\caption{Average beampattern matching error versus the SINR threshold $\Gamma$. }
	\label{fig:A-I-Rich}
\end{figure}

In the simulation, we consider half-wavelength spacing at the ULA. The power constraints of all the approaches are sum-power constraints with power budget $P_0 = 0.1$ Watt (W) or 20 dBm. The received signal at each communication user is corrupted with an additive white Gaussian noise of equal variance $\sigma^2 = -70$ dBm. Without loss of generality, we set $\Gamma_i = \Gamma, \forall i \in \mathcal{K}$. A pathloss of 80 dB is assumed between the BS and each communication receiver.
We also set the number of users $K=5$ and the number of transmit antennas $N = 8$ unless stated otherwise.
Without loss of generality, the desired beampattern is composed of five main beams, whose direction set $\Theta_d=\{-\frac{\pi}{3}, -\frac{\pi}{6}, 0, \frac{\pi}{6} ,\frac{\pi}{3}\}$. The width of each main beam $\delta = \frac{\pi}{18}$ and the grids are obtained by specifying $M=101$ discrete sensing angles ranging from $[-\frac{\pi}{2}:\frac{\pi}{100}:\frac{\pi}{2}]$. In conclusion, the normalized $\tilde{\mathcal{P}}(\theta)$ is defined as

\begin{align}
	\tilde{\mathcal{P}}(\theta)=\left\{\begin{array}{l}
		1, \theta_d-\frac{\delta}{2} \leq \theta \leq \theta_d+\frac{\delta}{2}, \theta_d \in \Theta_d \\
		0, \text { otherwise }
	\end{array}\right.
\end{align}

\subsection{SINR-Constrained Beampattern Matching} \label{Subsection-A}

This subsection considers the beampattern matching design. Figs. \ref{fig:A-I-Rich-a} and \ref{fig:A-I-Rich-b} show the beampattern matching error versus the SINR threshold $\Gamma$ in the case with Rayleigh fading channels and LOS channels, respectively, 
and the results are obtained by averaging over 200 random channel realizations. Unless stated otherwise, the direction of the users $\theta_i$ are randomly generated from the uniform distribution over $[-\frac{\pi}{2},\frac{\pi}{2}]$ in each LOS channel realization.
For performance comparison, we also consider the radar sensing only (i.e., beampattern matching in (P1) without communication users) as the sensing performance upper bound, and the two-stage design approach without dedicated radar signals in \cite[Section III-C]{liu2018mu} as a benchmark\footnote{Notice that in \cite{liu2018mu}, Gaussian randomization is needed to obtain a feasible but sub-optimal rank-one information beamforming solution. Here, we directly adopt the high-rank solution before the Gaussian randomization for the purpose of comparison only, which generally achieves better performance or a lower average matching error than that in \cite{liu2018mu}.}.

\begin{figure}[t]
	\centering
	\setlength{\abovecaptionskip}{-4mm}
	\setlength{\belowcaptionskip}{-4mm}
	\setlength{\abovecaptionskip}{+3pt}
	\subfigure[The case with Rayleigh fading channels]{ \label{fig:A-II-Tx-Beam-Rich}
		\includegraphics[width=2.6in]{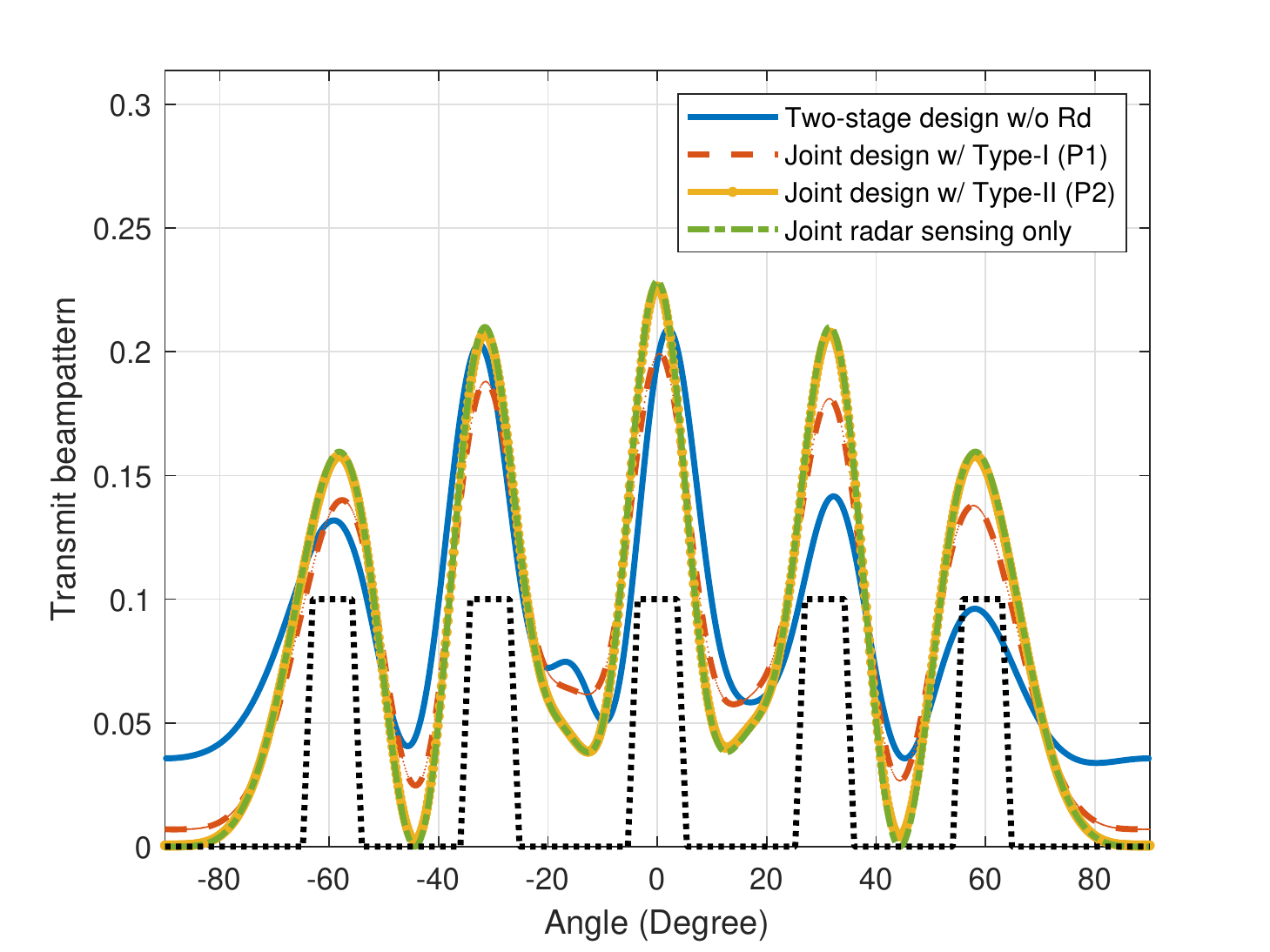}}
	\subfigure[The case with LOS channels]{ \label{fig:A-II-Tx-Beam-LOS}
		\includegraphics[width=2.6in]{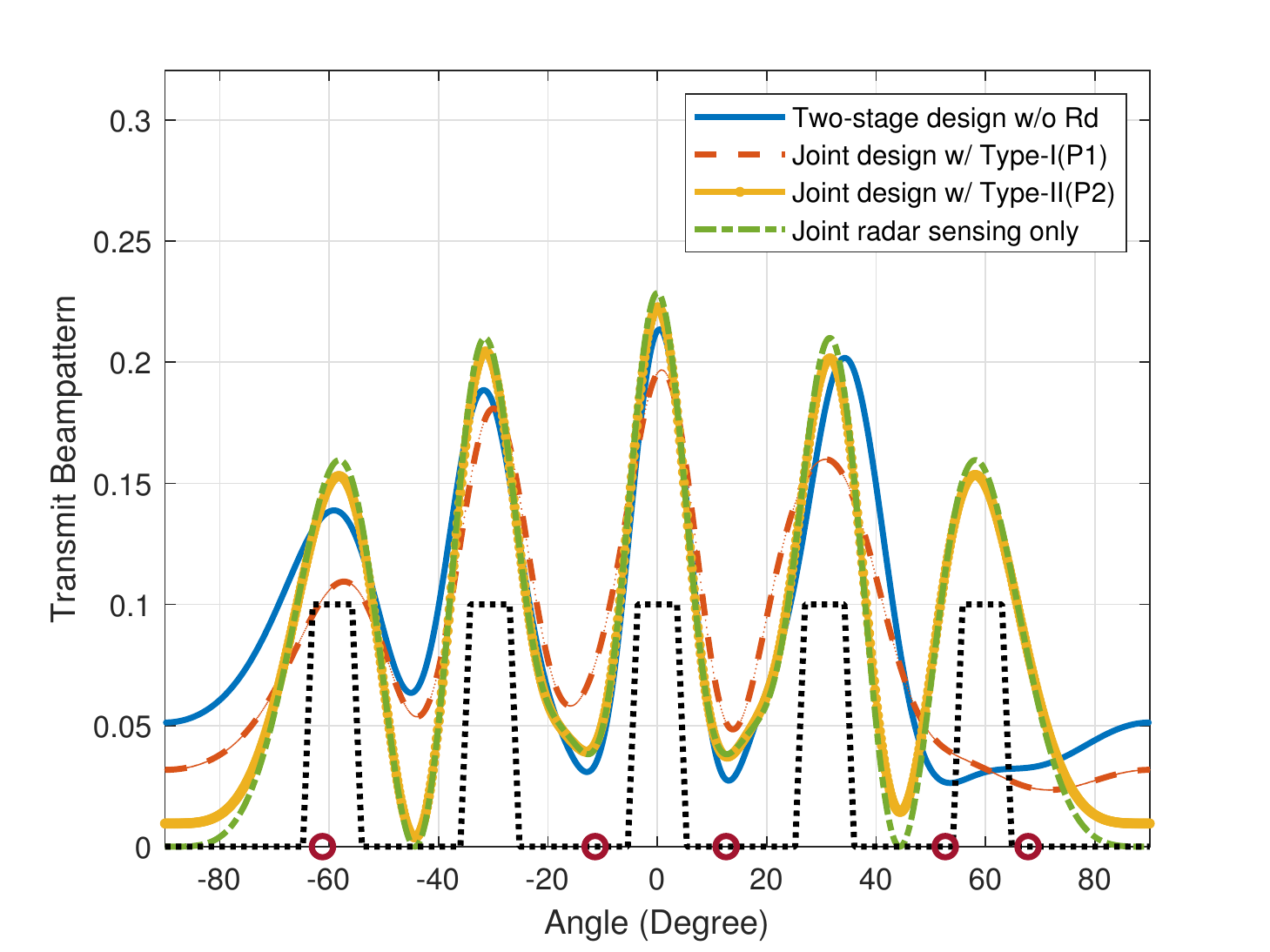}}
	\caption{Obtained transmit beampattern via the beampattern matching under Rayleigh fading channels and LOS channels with $\Gamma$ = 20 dB. The black dotted line specifies the desired transmit beampattern $\tilde{\mathcal{P}}(\theta)$. For LOS channels, the five red circles indicate the directions of five users.}
	\label{fig:A-II-Tx-Beam}
\end{figure}

In Fig. \ref{fig:A-I-Rich}, it is observed that the proposed joint design with Type-II receivers achieves much lower matching error than the other designs under both channel conditions.
It is also observed that under Rayleigh fading channels, the solution to (P1) satisfies that $\mv{R}_d = 0$ and $\text{rank}(\mv{T}_i) = 1, \forall i \in \mathcal{K}$ when $\Gamma \geq 6 $ dB and thus radar signal is not needed in this case under Type-I receivers. The result is not surprising, since when $\Gamma$ is high, the total transmit power should be used to form information signal beams to meet the SINR requirements at users, and no radar signals can be added as they may introduce harmful interference towards Type-I receivers. A similar phenomenon is observed in the case with LOS channels for Type-I receivers, which is consistent with Proposition \ref{T_30} and \ref{T_LOS}. 


Fig. \ref{fig:A-II-Tx-Beam} shows the obtained transmit sensing beampattern under $\Gamma$ = 20 dB by considering both Rayleigh fading and LOS channels. It is observed that the performance of our proposed design with Type-II receivers is close to the upper bound by the radar sensing only under both channel conditions. In particular, it is observed from Fig. \ref{fig:A-II-Tx-Beam-LOS} that when the channels are LOS, only the design with Type-II receivers achieves a beampattern similar to that by the radar sensing only, especially at the interested sensing region around $60^o$. This validates again the benefit of  adding dedicated radar signal together with the radar interference cancellation at the Type-II receivers in enhancing ISAC performance. 

\begin{figure}[t]
	\centering
	\setlength{\abovecaptionskip}{-4mm}
	\setlength{\belowcaptionskip}{-4mm}
	\setlength{\abovecaptionskip}{+3pt}
	\subfigure[The case with Type-I receivers]{ \label{fig:A-II-Tx-Beam-LOS-I-decomp}
		\includegraphics[width=2.6in]{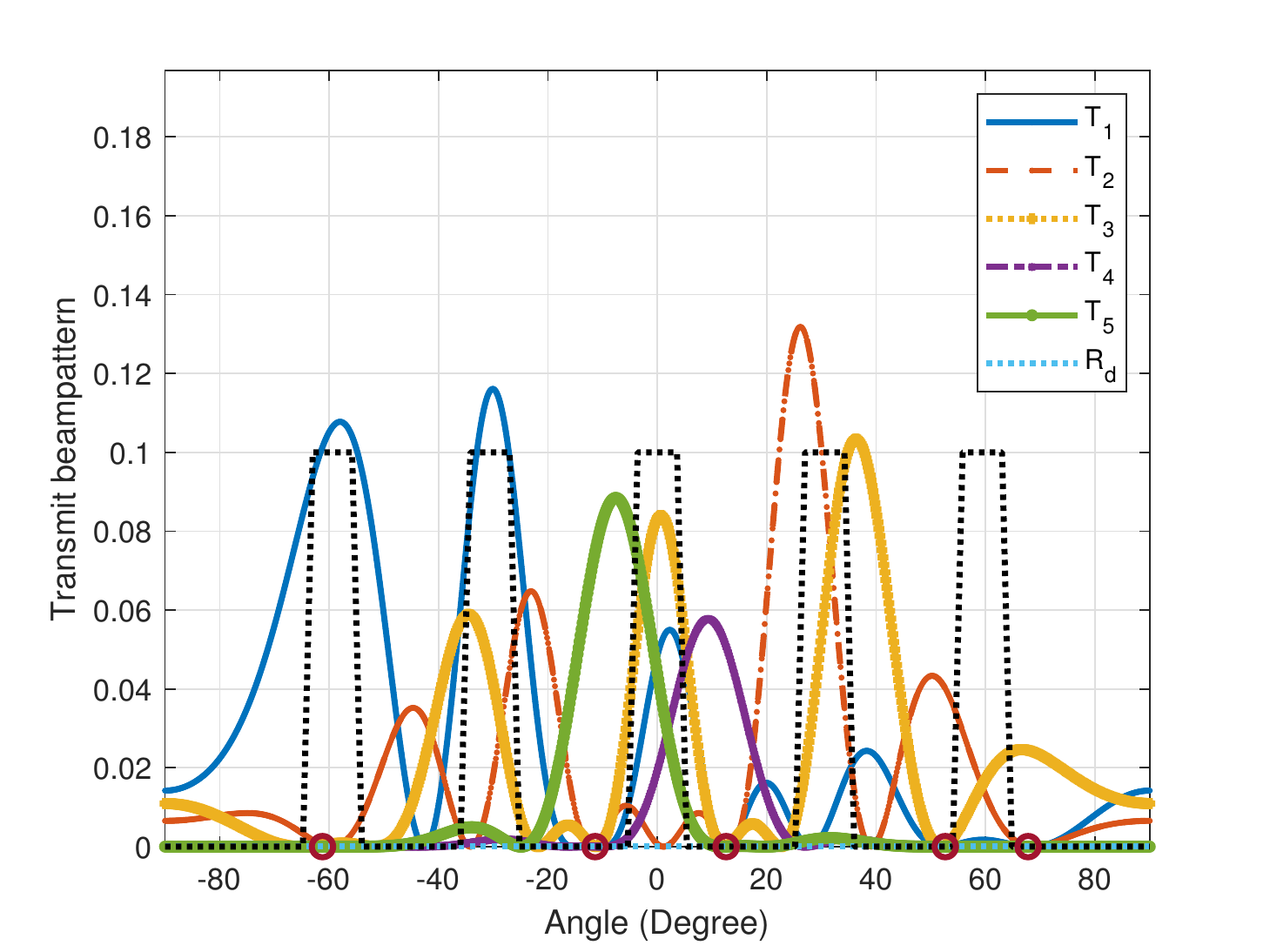}}
	\subfigure[The case with Type-II receivers]{ \label{fig:A-II-Tx-Beam-LOS-II-decomp}
		\includegraphics[width=2.6in]{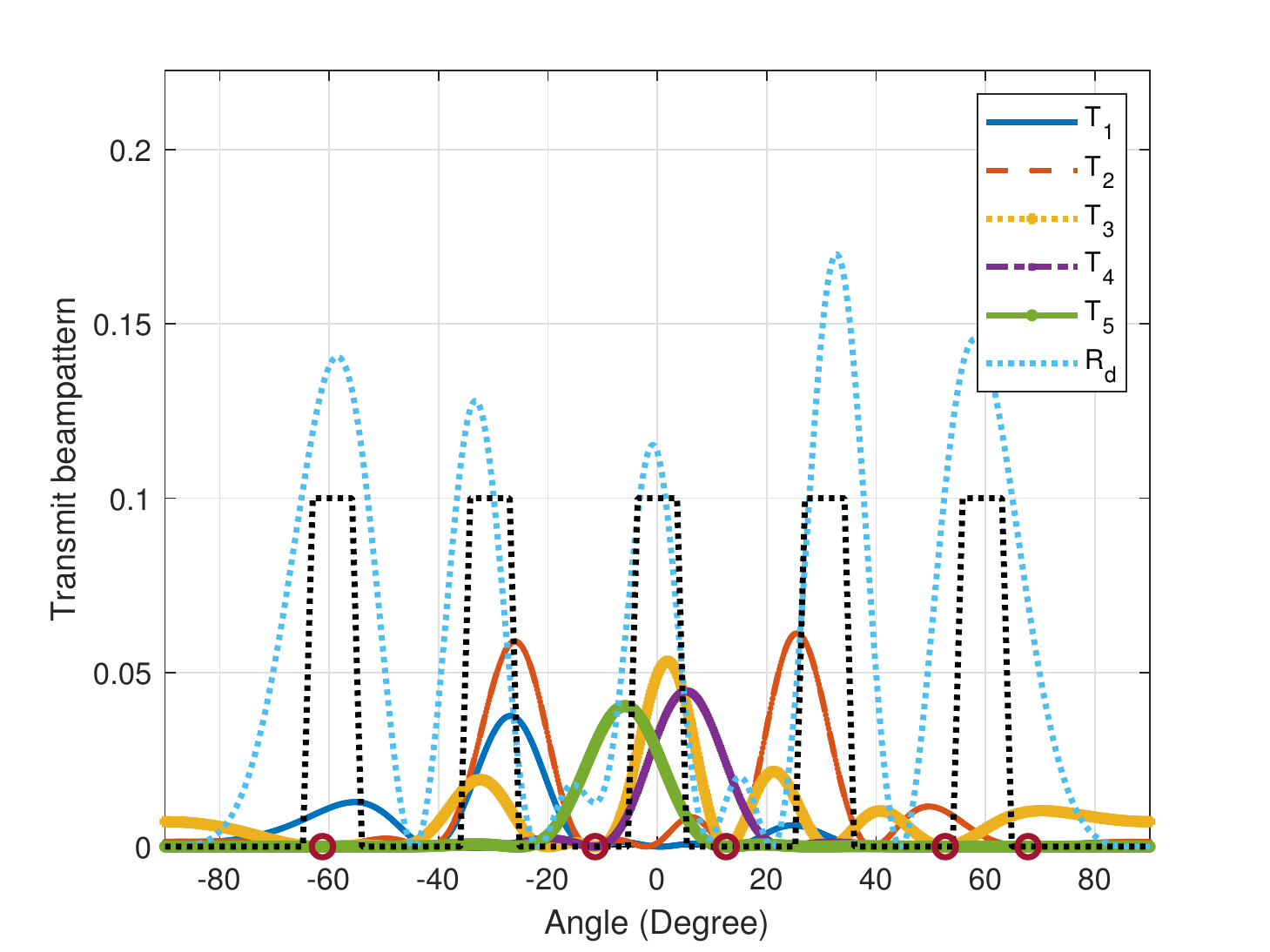}}
	\caption{Transmit beampattern achieved by individual information and radar signals, with Type-I and Type-II receivers under LOS channels. The black dotted line specifies the desired transmit beampattern $\tilde{\mathcal{P}}(\theta)$ while the five red circles indicate the directions of five users.}
	\label{fig:A-II-Tx-Beam-decomp}
\end{figure}
\begin{figure}[htb]
	\centering
	\setlength{\abovecaptionskip}{-4mm}
	\setlength{\belowcaptionskip}{-4mm}
	\setlength{\abovecaptionskip}{+3pt}
	\subfigure[The case with Rayleigh fading channels]{ \label{fig:B-I-Rich-a}
		\includegraphics[width=2.6in]{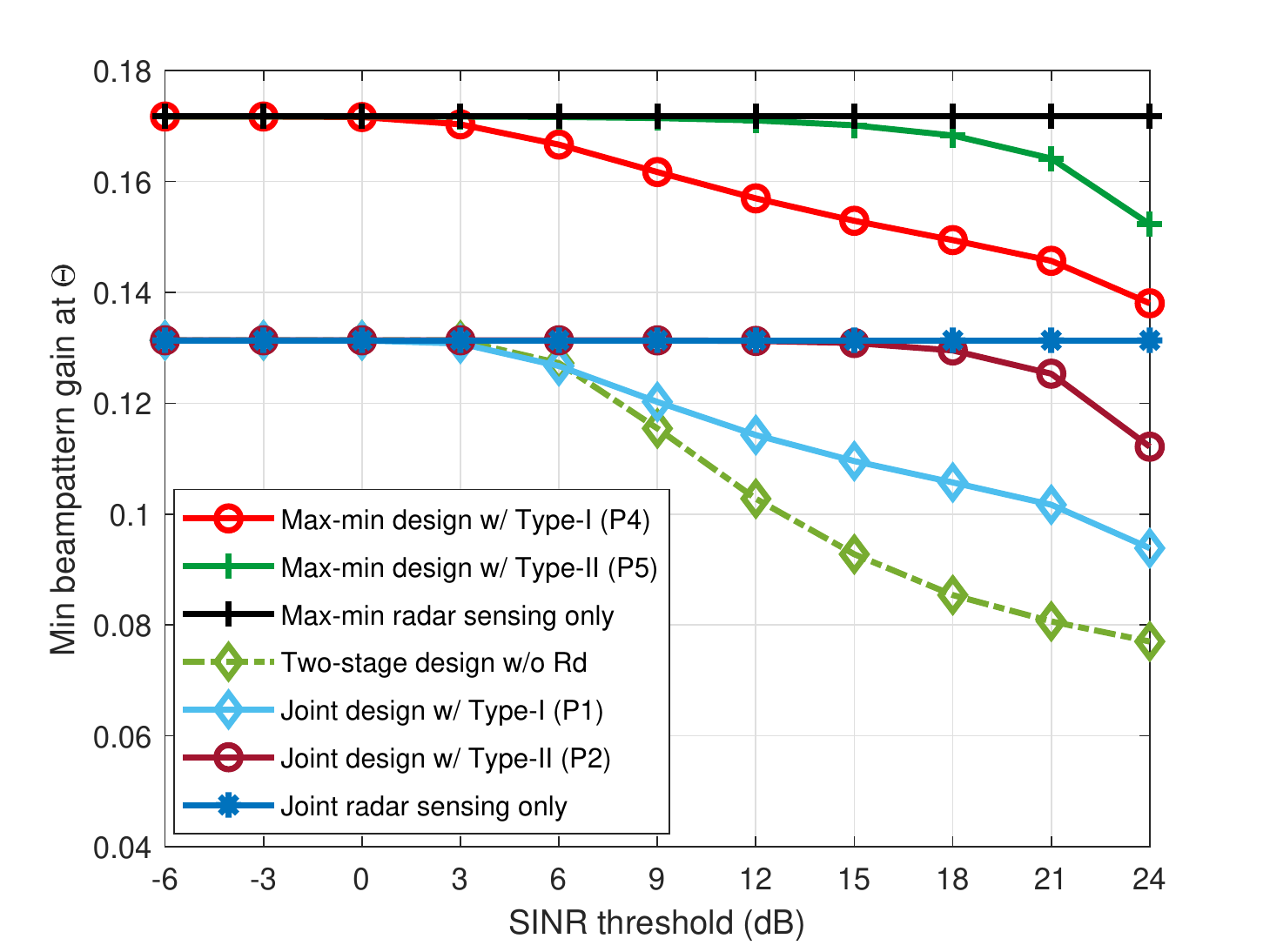}}
	\subfigure[The case with LOS channels]{ \label{fig:B-I-Rich-b}
		\includegraphics[width=2.6in]{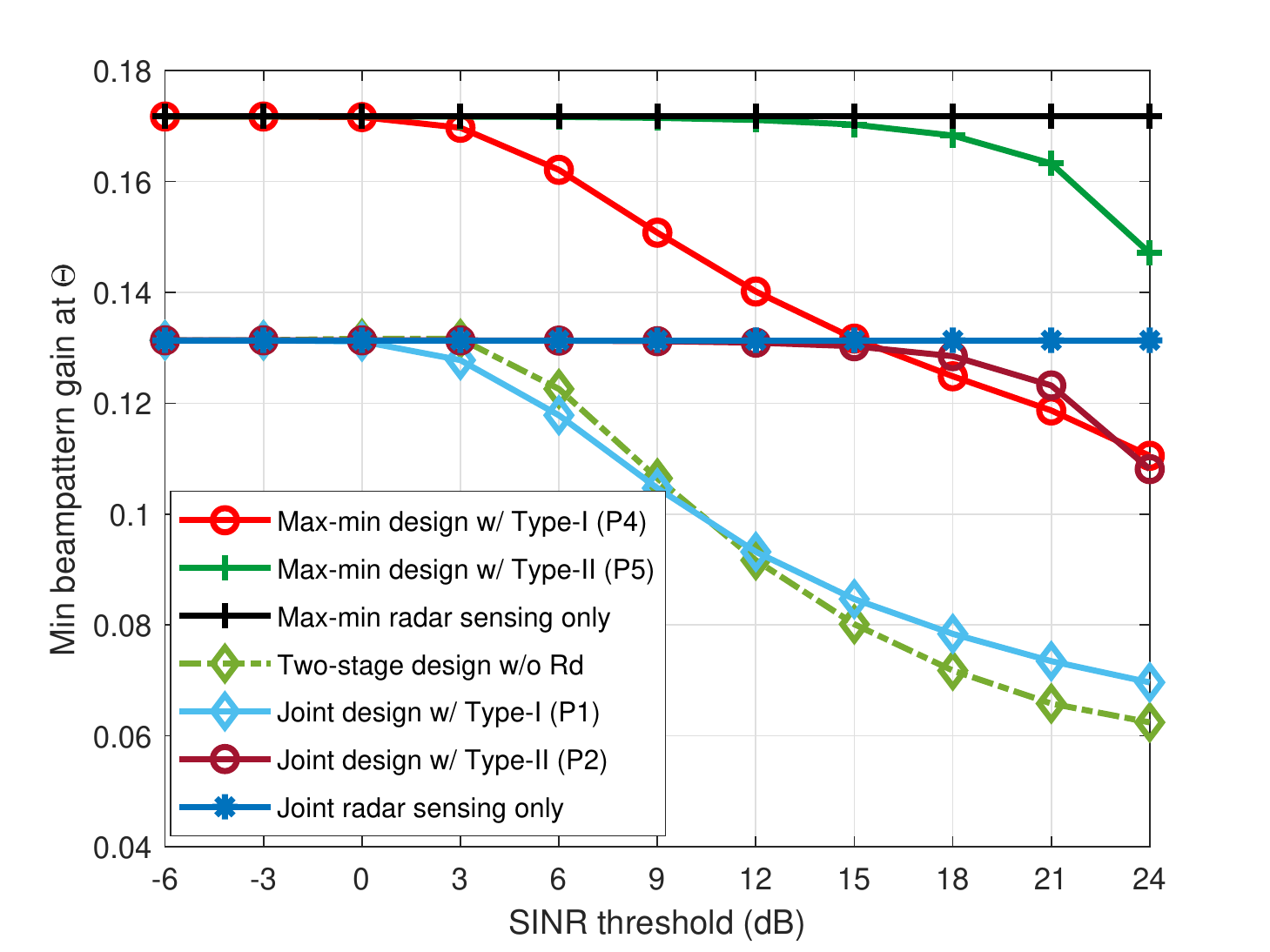}}
	\caption{The minimum weighted beampattern gain versus SINR threshold $\Gamma$. }
	\label{fig:B-I-Rich}
\end{figure}

To gain further insights,  Fig. \ref{fig:A-II-Tx-Beam-LOS-I-decomp} and Fig. \ref{fig:A-II-Tx-Beam-LOS-II-decomp} show the transmit beampatterns achieved by individual information and radar signals (i.e., $\mv{a}^H(\theta_m) \mv{T}_k \mv{a}(\theta_m), \forall k \in \mathcal{K}$ and $\mv{a}^H(\theta_m) \mv{R}_d \mv{a}(\theta_m)$) with Type-I and Type-II receivers, respectively, which are obtained by decomposing the achieved transmit beampattern in Fig. \ref{fig:A-II-Tx-Beam-LOS} under both types of receivers.
It is observed in Fig. \ref{fig:A-II-Tx-Beam-LOS-I-decomp} that with Type-I receivers, the transmit beampattern by the radar signal (labeled as $\bm{R}_d$) is zero to avoid the radar interference. In this case, the sensing requirements are ensured by reusing the information signals. By contrast, it is observed in Fig. \ref{fig:A-II-Tx-Beam-LOS-II-decomp} that with Type-II receivers, the radar signal results in large transmit beampatterns in interested sensing regions to maintain the sensing performance, which will not degrade the communication SINR, as the resultant interference can be canceled at these communication receivers. As such, the beampatterns by the communication signals (labeled as $\bm{T}_k$'s, $k \in \{1,...,5\}$) in Fig. \ref{fig:A-II-Tx-Beam-LOS-II-decomp} are observed to be much smaller than those in Fig. \ref{fig:A-II-Tx-Beam-LOS-I-decomp}. In particular, in Fig. \ref{fig:A-II-Tx-Beam-LOS-II-decomp} with Type-II receivers,
the sensing requirements around $-60^o$ and $-30^o$ are observed to be ensured mainly by the radar signal, while in Fig. \ref{fig:A-II-Tx-Beam-LOS-I-decomp} with Type-I receivers, they are ensured by the communication signal for receiver 1 ($\bm{T}_1$).
Furthermore, it is observed in Fig. \ref{fig:A-II-Tx-Beam-LOS-I-decomp} that under our considered setup, when the users are located in the sensing regions (i.e., user around $-60^o$), the corresponding beampattern gain by $\bm{T}_1$ is much higher than those by $\bm{T}_k, k \in \{2,...,5\}$, for the users outside the sensing regions (i.e., users around $-10^o, 10^o, 55^o$, and $65^o$).
The reason is that for the sensing, any beampattern gain outside the sensing regions is not desired while for the communication, certain beampattern gain is still required to meet the SINR requirement of the users outside the sensing regions. This inherent trade-off outside the sensing region will result in a relatively smaller beampattern gain for users around $-10^o$ and $10^o$, compared with that for the user around $-60^o$ in the sensing region.

\subsection{SINR-Constrained Minimum Weighted Beampattern Gain Maximization} \label{Subsection-B}

In this subsection, we consider the designs with minimum weighted beampattern gain maximization (abbreviated as ‘‘max-min design’’). Instead of considering the specified $M$ point angles in Section \ref{Subsection-A},  here we only need to choose a subset of them (with positive desired beampattern gains) as the angles of interest $\Theta = \{ \theta_1, \theta_2, ..., \theta_Q\}$. In particular, in the simulation, $\Theta$ is composed of the angles with non-zero desired beampattern gains in Fig. \ref{fig:A-II-Tx-Beam}. Without loss of generality, we set $\eta_\theta = 1, \forall \theta \in \Theta$.

Figs. \ref{fig:B-I-Rich-a} and \ref{fig:B-I-Rich-b} show the obtained minimum weighted beampattern gain at $\Theta$ versus the SINR threshold $\Gamma$ by considering the Rayleigh fading and LOS channels, respectively.
The results are obtained by averaging over 200 random channel realizations. It is observed that the max-min designs achieve much higher worst-case beampattern gains at angles of interest than the beampattern matching designs. In particular, under both design criteria, the case with Type-II receivers is observed to have much better performance than other schemes, 
especially when the SINR threshold becomes high (e.g., $\Gamma \geq 10 $ dB), 
thanks to its capability of cancelling the interference caused by radar signals.
\begin{figure}[t]
	\centering
	\setlength{\abovecaptionskip}{-4mm}
	\setlength{\belowcaptionskip}{-4mm}
	\setlength{\abovecaptionskip}{+3pt}
	\subfigure[The case with Rayleigh fading channels]{ \label{fig:B-III-Rich-a}
		\includegraphics[width=2.6in]{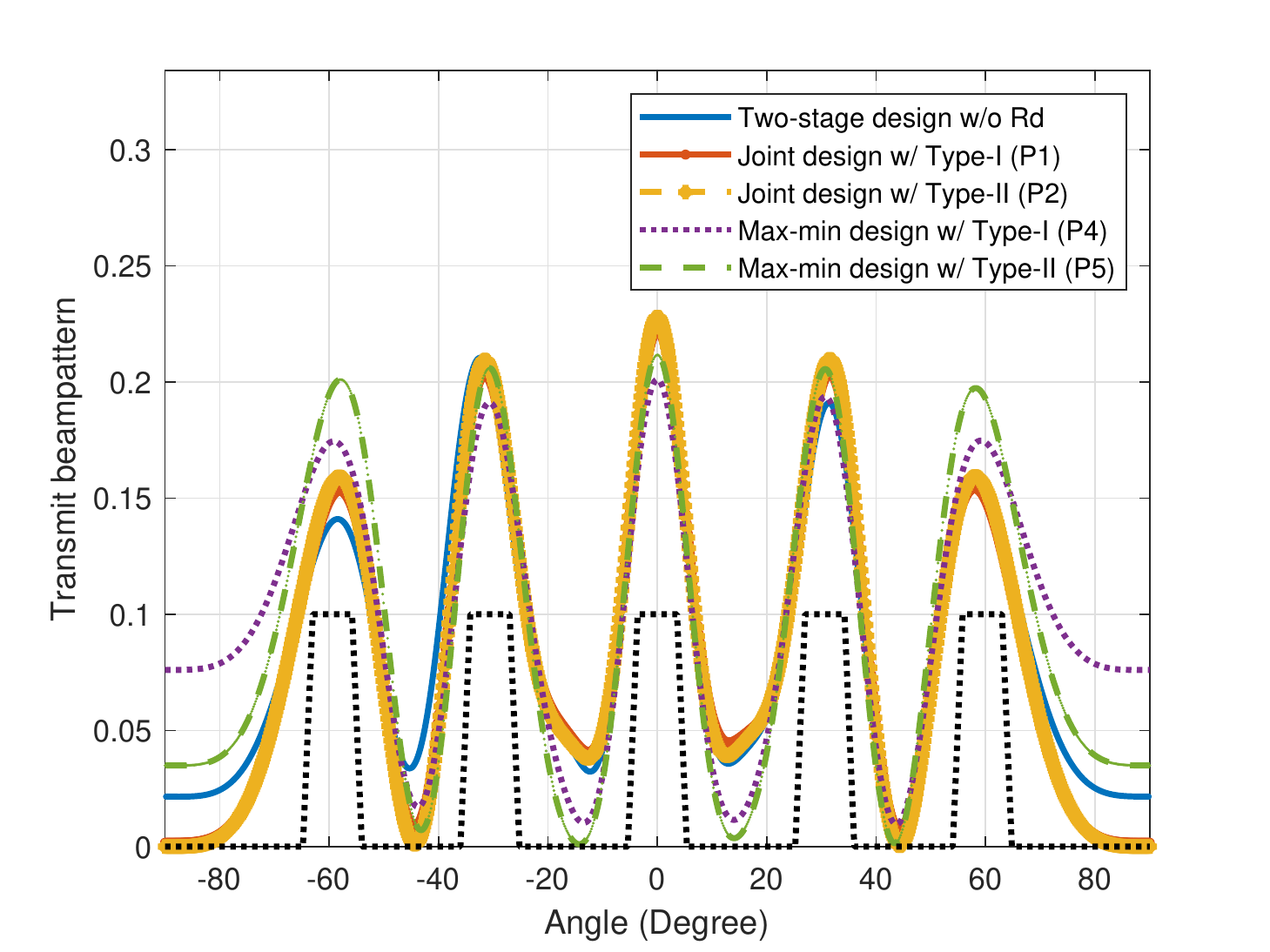}}
	\subfigure[The case with LOS channels]{ \label{fig:B-III-Rich-b}
		\includegraphics[width=2.6in]{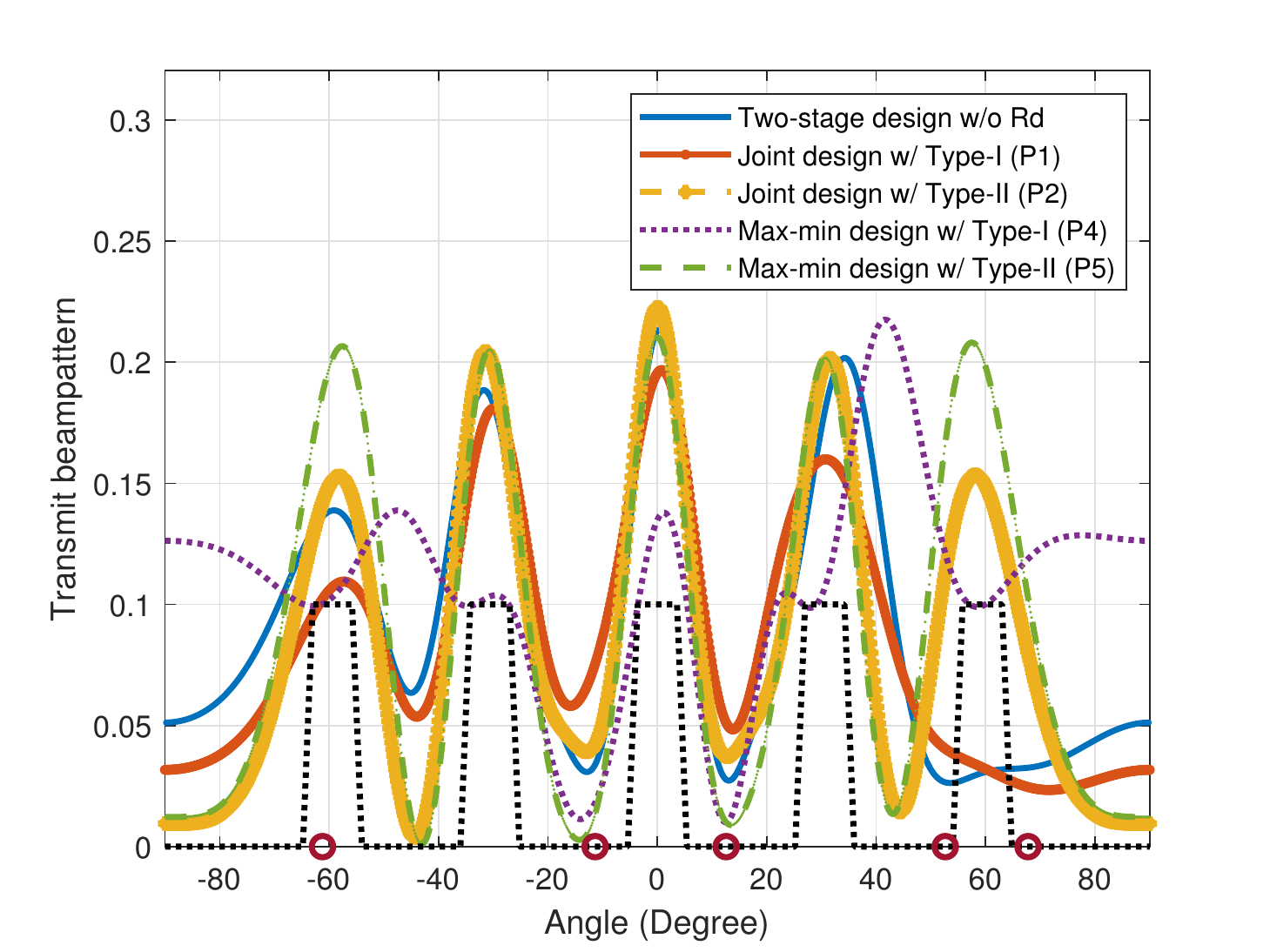}}
	\caption{Obtained transmit beampattern by both design criteria under both channel conditions with $\Gamma$ = 20 dB. The black dotted line specifies $\tilde{\mathcal{P}}(\theta)$. For LOS channels, the five red circles indicate the directions of five users.}
	\label{fig:B-III-Rich}
\end{figure}

Fig. \ref{fig:B-III-Rich} shows the transmit beampattern by the beampattern matching and max-min designs under both types of receivers. It is observed that the max-min design with Type-II receivers (P5) yields more balanced beampatterns in the interested sensing regions compared with other designs under both channel conditions. Besides, the max-min design with Type-I receivers (P4) is observed to have distorted beampatterns and degraded performance compared with the case with Type-II receivers (P5), again validating the benefits brought by adding dedicated radar signal together with the radar interference cancellation. In particular, under LOS channels in Fig. \ref{fig:B-III-Rich-b}, the beampattern via (P4) is observed to be significantly distorted, as compared to that via (P5) to meet the SINR requirement. Nevertheless, the beampattern via the max-min design (P4) is still observed to outperform that by the beampattern matching design (P1). In particular, the resultant beampattern gain via (P4) is observed to be more uniformly distributed at all the interested sensing regions, with the minimum value higher than 0.1. By contrast, the resultant beampattern gain via (P1) is observed to be less than 0.04 at the sensing region around $60^o$. This thus validates the effectiveness of the max-min beampattern gain designs.
\begin{figure}[t]
	\centering
	\setlength{\abovecaptionskip}{-4mm}
	\setlength{\belowcaptionskip}{-4mm}
	\setlength{\abovecaptionskip}{+3pt}
	\subfigure[The case with Rayleigh fading channels]{ \label{fig:B-IV-Rich-a}
		\includegraphics[width=2.6in]{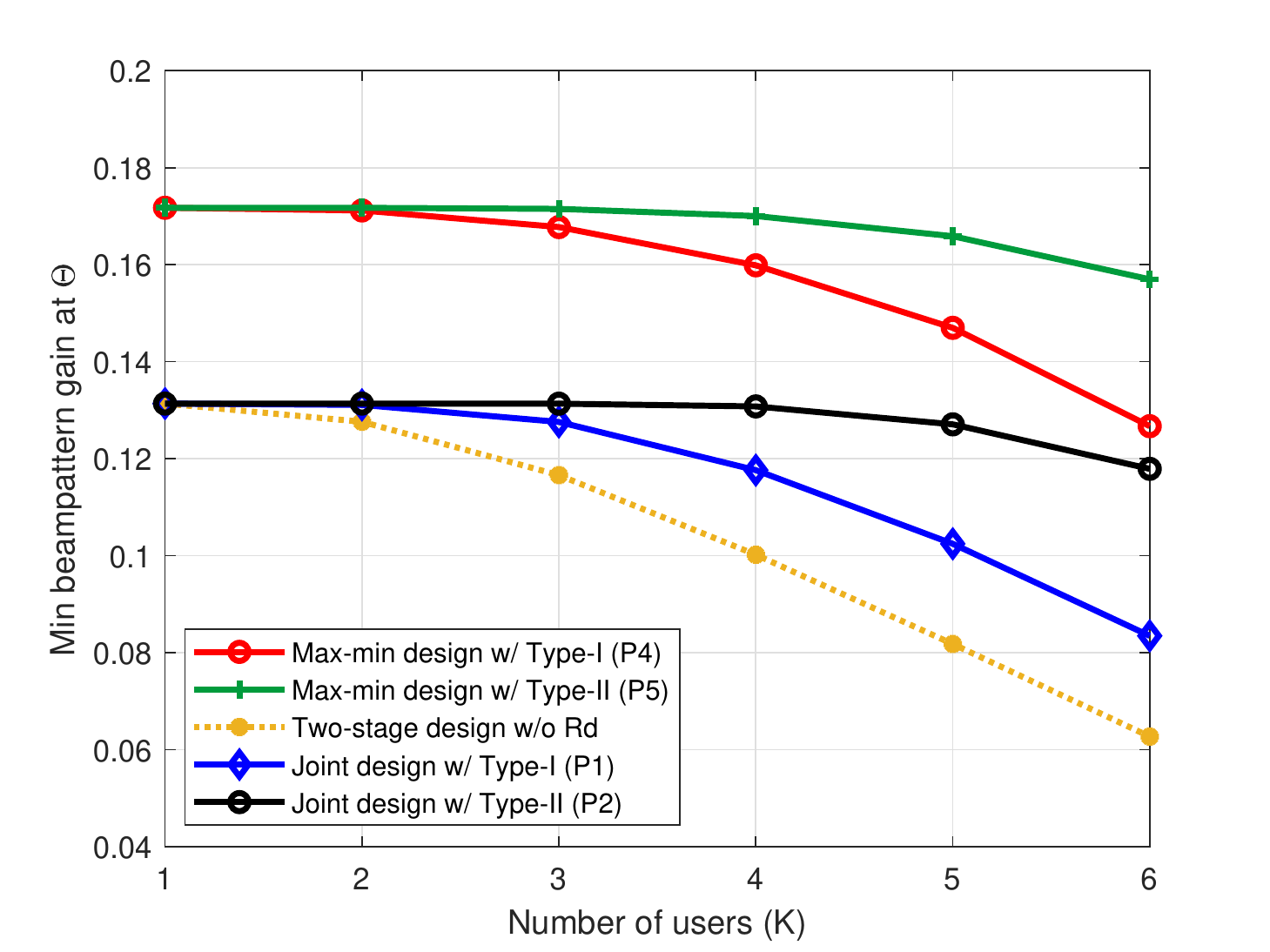}}
	\subfigure[The case with LOS channels]{ \label{fig:B-IV-Rich-b}
		\includegraphics[width=2.6in]{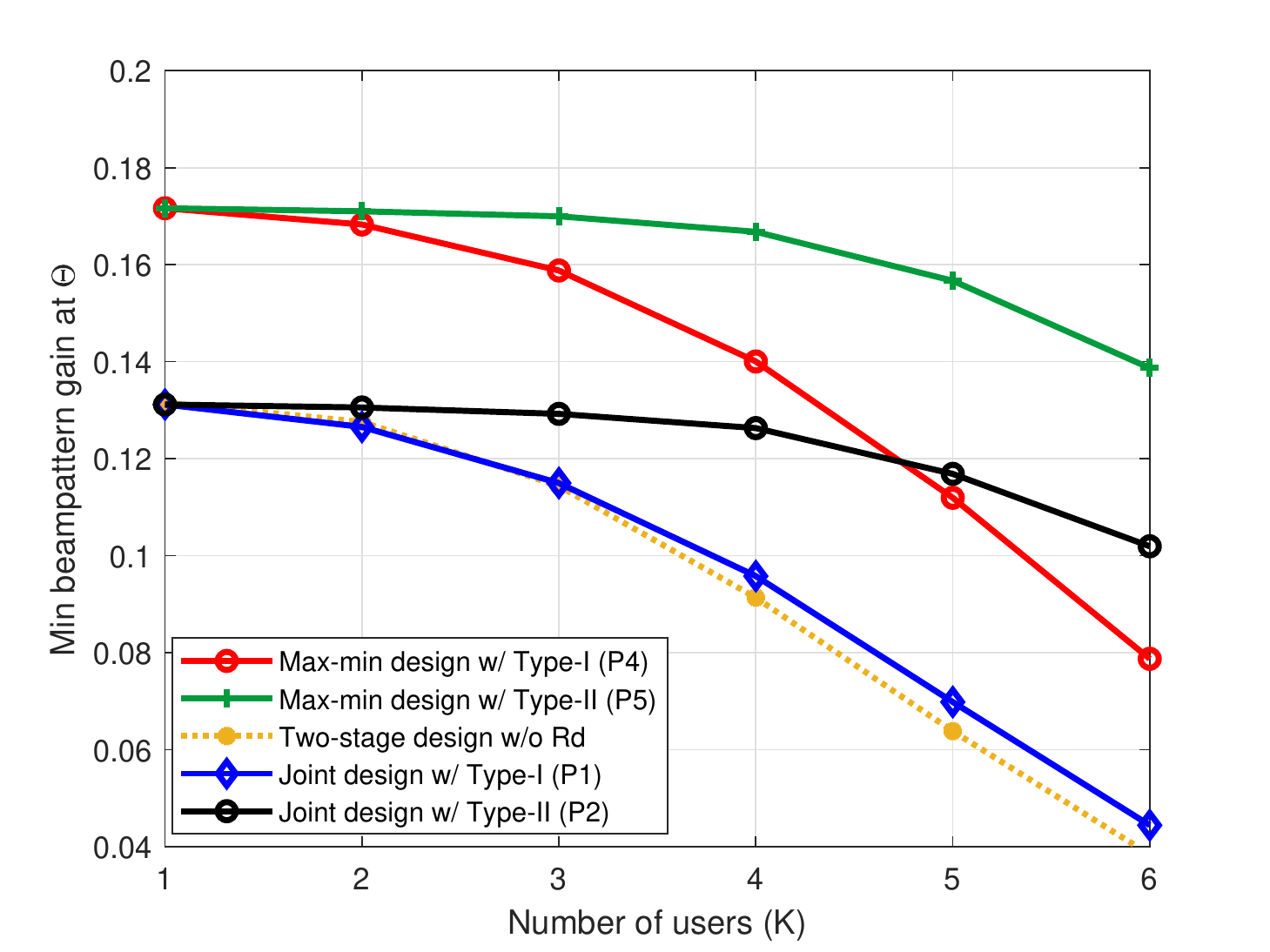}}
	\caption{Minimum weighted beampattern gain at angles of interest $\Theta$ obtained by five different methods with $\Gamma$ = 20 dB.} \label{fig:B-IV-Rich}
\end{figure}

Fig. \ref{fig:B-IV-Rich} shows the minimum weighted beampattern gain at $\Theta$ versus the number of users $K$, where we set $\Gamma = 20$ dB and the number of antennas as $N=8$. The results in Figs. \ref{fig:B-IV-Rich-a} and \ref{fig:B-IV-Rich-b} are obtained by averaging over 200 Rayleigh and LOS channel realizations, respectively. As number of users increases, the minimum weighted beampattern gain at $\Theta$ are observed to decrease. The proposed max-min design with Type-II receivers is observed to outperform all other approaches under both channel conditions.

\begin{figure}[t]
	\centering
	\setlength{\abovecaptionskip}{-4mm}
	\setlength{\belowcaptionskip}{-4mm}
	\setlength{\abovecaptionskip}{+3pt}
	\subfigure[The case with respect to $N$]{ \label{fig:C-III-Rich-a}
		\includegraphics[width=2.6in]{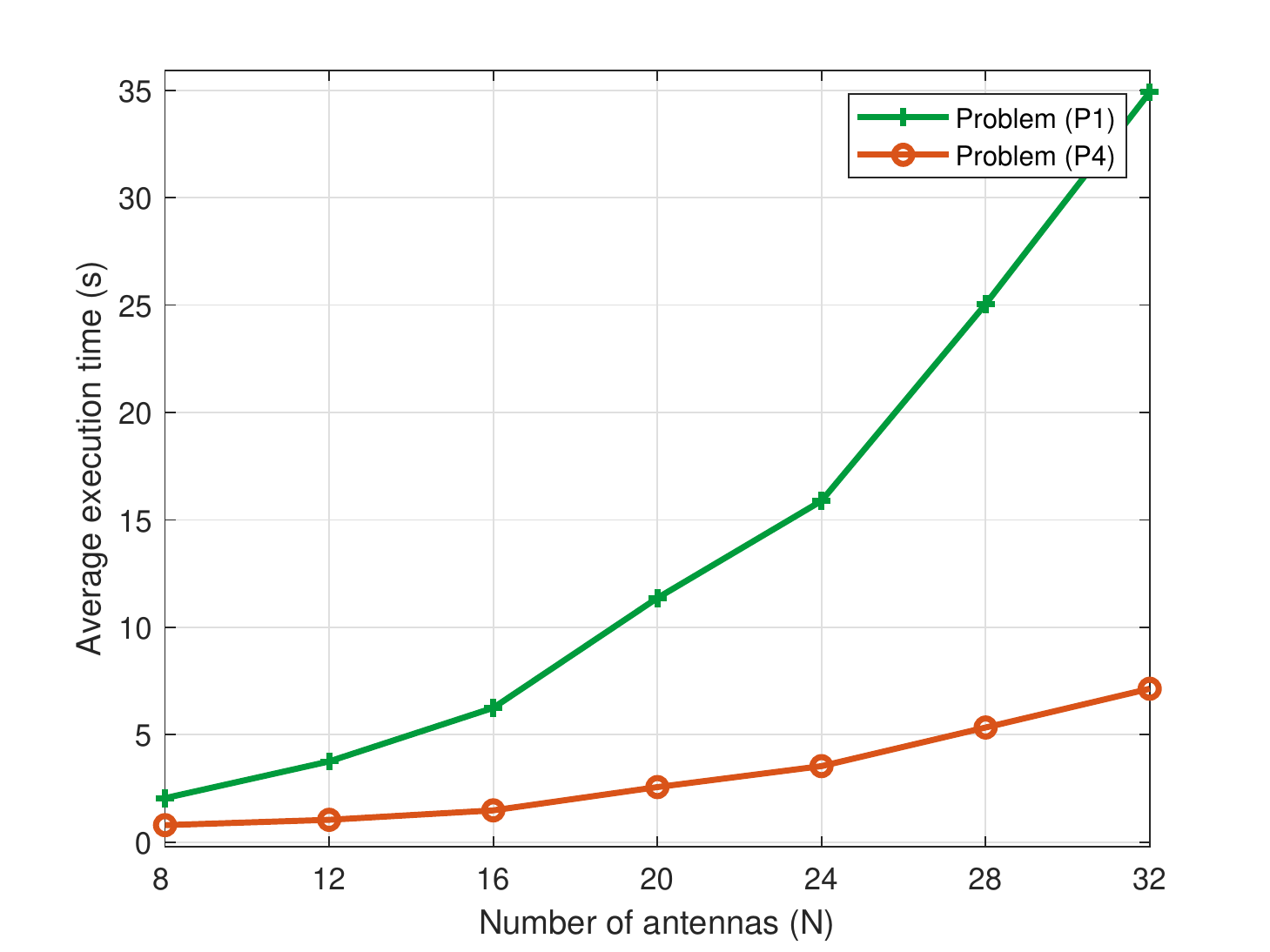}}
	\subfigure[The case with respect to $K$]{ \label{fig:C-III-Rich-b}
		\includegraphics[width=2.6in]{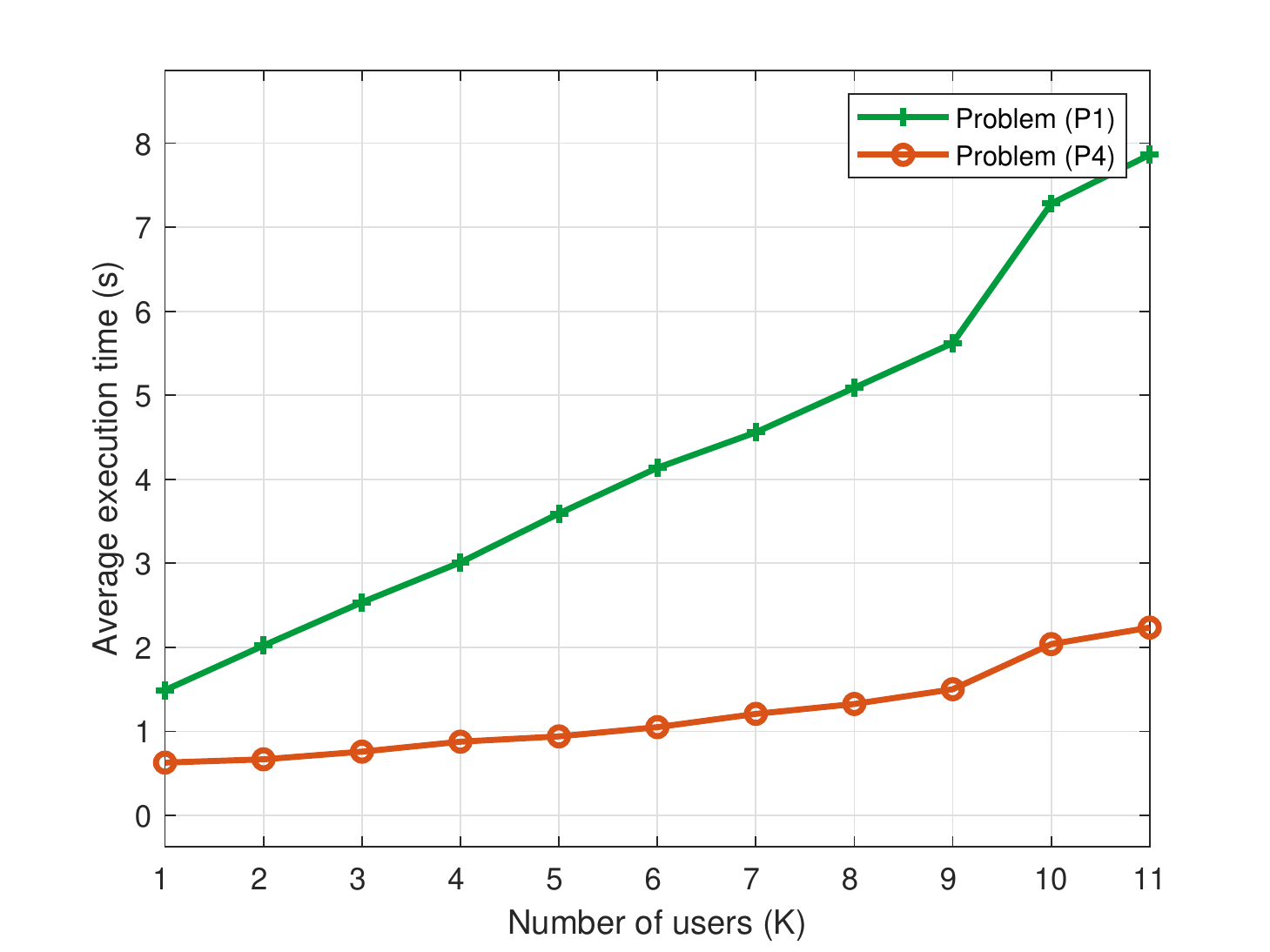}}
	\caption{The execution time versus the number of antennas $N$ and the number of users $K$.} \label{fig:C-III-Rich}
\end{figure}

Finally, Fig. \ref{fig:C-III-Rich} compares the computational complexity in terms of execution time between the aforementioned two designs (P4) versus (P1) under Rayleigh fading channels, which is obtained by averaging over 200 random realizations. 
In Fig. \ref{fig:C-III-Rich-a}, we set $\Gamma= 20$ dB and $K=5$ and show the execution time with respect to number of antennas $N$; while in Fig. \ref{fig:C-III-Rich-b}, we set $N=12$, $\Gamma = 20$ dB and show the execution time with respect to number of users $K$. Both figures show that the execution time of solving (P1) grows much faster than that of solving (P4), indicating the benefit of the max-min design in terms of the implementation complexity. This is consistent with the analysis in Section \ref{Section_complexity}.

\section{Concluding Remarks} \label{Section_conclusion}

This paper considered a downlink ISAC system, in which one BS equipped with ULA jointly designs the information and radar transmit beamforming to perform downlink multiuser communication and target sensing simultaneously. We studied two design criteria - the beampattern matching and the newly proposed one aiming to maximize the minimum weighted beampattern gain. By considering the addition of dedicated radar signals, we introduced two types of communication receivers (Type-I and Type-II) without and with the capability of radar interference cancellation. Though non-convex in general, we optimally solved the two design problems under both types of receivers, by applying the technique of SDR.  Under both design criteria, it is rigorously proved that equipping the ISAC system with dedicated radar signal in general enhances the performance under both types of receivers, and Type-II receivers yield better performance compared with Type-I receivers. Furthermore, when the wireless channels are LOS, dedicated radar signal is not needed with Type-I receivers. It was shown that the minimum weighted beampattern gain maximization design with Type-II receivers achieves the best joint sensing and communication performance with significantly lower computational complexity. It is expected that the design principles in this paper can also be extended to other setups such as those with intelligent reflecting surface (IRS), hybrid beamforming, wideband transmission, and multi-node cooperation. In the following, we briefly discuss the possible extensions to motivate future research. 
\begin{itemize}
	\item Recently, IRS or reconfigurable intelligent surface (RIS) has been recognized as a promising technique to enhance the communication performances by reconfiguring the radio propagation environments \cite{ris-ISAC-joint}, and also improve the sensing performance by providing additional sensing paths \cite{shao2022target, song2022intelligent, song2022joint}. For such IRS-enabled ISAC systems, how to optimize the sensing performance (e.g., beampattern gains) while ensuring the communication SINR requirements is an interesting new problem, for which the reflective beamforming at the IRS becomes a new design degree of freedom for optimization, in addition to the transmit beamforming at the BS.
	\item While this paper focused on the fully digital transmitter architecture at the BS for ISAC, the hybrid analog-digital architecture is becoming more and more popular with the advancements of massive MIMO and mmWave/THz communications. In this case, designing proper hybrid architectures and optimizing the corresponding hybrid transmit beamforming are new problems for efficient ISAC. However, finding the optimal solution to such problems can be quite challenging due to, e.g., the coupling between analog and digital beamformers and the unit modulus constraints for analog beamforming. 
	\item This paper studied the multi-antenna ISAC over narrowband channels. It is an interesting topic to extend our proposed designs to wideband systems. In general, wideband signals can be exploited for sensing to extract more target information such as Doppler and range, but how to design the transmit beamforming over different subcarriers is challenging. In particular, with wideband transmission, choosing the appropriate sensing performance metrics is crucial. Different from the communication that might adopt individual SINR over each subcarrier as the performance measure, the sensing performance is jointly determined by the echo signals over all subcarriers. More research along this direction is needed to make ISAC more practical and compatible with current technologies.
	\item Last but not least, this work considered a single BS or ISAC transceiver. Enabling multiple BSs or ISAC transceivers to collaborate for networked ISAC \cite{huangyi_network,rahman2019framework} is important to enhancing the sensing and communication performances. On one hand, different BSs can employ the coordinated or joint transmit beamforming for better managing the inter-cell interference for communications. On the other hand, these BSs can sense targets from different perspectives, and aggregate the received information at central controllers (e.g., central cloud) for cooperative sensing and localization. New sensing performance measures beyond the transmit beampatterns are needed for the networked ISAC designs. 
\end{itemize} 

\appendix


\subsection{Proof of Proposition \ref{T_2}}\label{Proof_theorem P2}
Suppose that $\{\{{\hat{\mv{T}}_k}\}, \hat{\mv{R}}_d, \hat{\alpha} \}$ denotes an optimal solution to (SDR2), where $\text{rank}(\hat{\mv{T}}_k) \ge 1, \forall k \in \mathcal{K}.$ In this case, we construct a new solution $\{\{\bar{\mv{T}}_k\},\bar{\mv{R}}_d,\bar{\alpha}\}$, given by
\begin{subequations}
	\begin{align}
		\label{equ:alpha_1}
		&\bar{\alpha} = \hat{\alpha}, \\
		\label{equ:Tk_1}
		&\bar{\mv{t}}_k = (\mv{h}_k^H \hat{\mv{T}}_k \mv{h}_k)^{-1/2} \hat{\mv{T}}_k \mv{h}_k, \bar{\mv{T}}_k = \bar{\mv{t}}_k \bar{\mv{t}}_k^H, \\
		\label{equ:Rd_1}
		&\bar{\mv{R}}_d = \sum_{k=1}^K \hat{\mv{T}}_k + \hat{\mv{R}}_d - \sum_{k=1}^K \bar{\mv{T}}_k, \forall k \in \mathcal{K}.
	\end{align}
\end{subequations}
It follows from (\ref{equ:Tk_1}) that $\{\bar{\mv{T}}_k\}$ are all rank-one and positive semidefinite. According to (\ref{equ:alpha_1}) and (\ref{equ:Rd_1}), the objective value achieved by $\{\{\bar{\mv{T}}_k\},\bar{\mv{R}}_d,\bar{\alpha}\}$ remains the same and the corresponding power constraint is met with equality.

Then, for any $\mv{v} \in \mathbb{C}^M$, it holds that for $\forall k \in \mathcal{K}$,
\begin{align}
	\mv{v}^H \left( \hat{\mv{T}}_k - \bar{\mv{T}}_k \right) \mv{v} = \mv{v}^H \hat{\mv{T}}_k \mv{v} -  \left(\mv{h}_{k}^{H} \hat{\mv{T}}_{k} \mv{h}_{k}\right)^{-1}  \left| \mv{v}_{k}^{H} \hat{\mv{T}}_{k} \mv{h}_{k} \right|^2.
\end{align}
According to the Cauchy-Schwarz inequality, we have
\begin{align}
	\left( \mv{v}^H \hat{\mv{T}}_k \mv{v} \right) \left(\mv{h}_{k}^{H} \hat{\mv{T}}_{k} \mv{h}_{k}\right) \geq \left| \mv{v}_{k}^{H} \hat{\mv{T}}_{k} \mv{h}_{k} \right|^2, \forall k \in \mathcal{K}.
\end{align}
Thus, we have
\begin{align} \label{PSD_subtract}
	\mv{v}^H \left( \hat{\mv{T}}_k - \bar{\mv{T}}_k \right) \mv{v} \geq 0  \Leftrightarrow  \hat{\mv{T}}_k - \bar{\mv{T}}_k \succeq 0, \forall k \in \mathcal{K}.
\end{align}
From (\ref{equ:Rd_1}), since $\hat{\mv{R}}_d \succeq 0$ and the summation of a set of positive semidefinite matrices is also positive semidefinite, it follows that $\bar{\mv{R}}_d \succeq 0$.

Furthermore, it can be easily shown that
\begin{equation} \label{Same_trans}
	\mv{h}_{k}^{H} \bar{\mv{T}}_{k} \mv{h}_{k}=\mv{h}_{k}^{H} \bar{\mv{t}}_{k} \bar{\mv{t}}_{k}^{H} \mv{h}_{k}=\mv{h}_{k}^{H} \hat{\mv{T}}_{k} \mv{h}_{k}, \forall k \in \mathcal{K}.
\end{equation}
Notice that the SINR constraints in (P2.1) can be reformulated as
\begin{align} \label{SINR_reform}
	\left(1+\frac{1}{\Gamma_{i}}\right) \bm{h}_{i}^H \bm{T}_{i} \bm{h}_{i}-  \bm{h}_{i}^H \left(\sum\limits_{k=1}^K \bm{T}_{k} \right) \bm{h}_{i} -\sigma_{i}^{2} \geq 0, \forall i \in \mathcal{K}.
\end{align}
Thus, we have
\begin{align}
	\nonumber
	& \left(1+\frac{1}{\Gamma_{i}}\right) \mv{h}_{i}^{H} \bar{\mv{T}}_{i} \mv{h}_{i}=\left(1+\frac{1}{\Gamma_{i}}\right) \mv{h}_{i}^{H} \hat{\mv{T}}_{i} \mv{h}_{i} \\
	\label{A}
	&\geq \mv{h}_{i}^{H} \left(\sum\limits_{k=1}^K \hat{\bm{T}}_{k} \right) \mv{h}_{i}+\sigma_i^{2} \geq \mv{h}_{i}^{H} \left(\sum\limits_{k=1}^K \bar{\bm{T}}_{k} \right)  \mv{h}_{i}+\sigma_i^{2}.
\end{align}
where the first equality follows from (\ref{Same_trans}), and the first and the second inequalities follow from (\ref{SINR_reform}) and (\ref{PSD_subtract}), respectively. As a result, $\{\bar{\mv{T}}_k\}$ and $\bar{\mv{R}}_d$ also ensure the SINR constraints at communication users. By combining the results above, Proposition 2 is finally proved.

\subsection{Proof of Proposition \ref{T_30}}\label{Proof_theorem P3}
First, we show that $\{\{\hat{\mv{T}}_k\},\hat{\alpha}\}$ is indeed a feasible solution to (SDR3). Since $\{\{\tilde{\mv{T}}_k\},\tilde{\mv{R}}_d,\tilde{\alpha}\}$ is the optimal solution to (SDR1), we have
\begin{align}
	\nonumber
	& \left(1+\frac{1}{\Gamma_{i}}\right) \mv{h}_{i}^{H} \hat{\mv{T}}_i \mv{h}_{i} = \left(1+\frac{1}{\Gamma_{i}}\right) \mv{h}_{i}^{H} \left( \tilde{\mv{T}}_i + \beta_i \tilde{\mv{R}}_d \right) \mv{h}_{i}\\
	\nonumber
	&\stackrel{(a)}{\geq} \left(1+\frac{1}{\Gamma_{i}}\right) \mv{h}_{i}^{H} \tilde{\mv{T}}_{i} \mv{h}_{i} \stackrel{(b)}{\geq} \mv{h}_{i}^{H} \left(\sum\limits_{k=1}^K \tilde{\bm{T}}_{k} + \tilde{\mv{R}}_d\right) \mv{h}_{i}+\sigma_i^{2} \\
	&
	=\mv{h}_{i}^{H} \left(\sum\limits_{k=1}^K \hat{\bm{T}}_{k} \right)  \mv{h}_{i}+\sigma_i^{2}, \forall i \in \mathcal{K},
\end{align}
where (a) follows due to the fact that $\beta_i \tilde{\mv{R}}_d \succeq  0$ and (b) holds as $\{\{\tilde{\mv{T}}_k\},\tilde{\mv{R}}_d,\tilde{\alpha}\}$ are feasible for (SDR1).
Besides, it follows that
\begin{align}
	\text{tr}\left( \sum_{k=1}^{K} \tilde{\bm{T}}_{k} + \tilde{\mv{R}}_d \right) = \text{tr}\left( \sum_{k=1}^{K} \hat{\bm{T}}_{k} \right) = P_0,
\end{align}
and it is easy to see that $\hat{\mv{T}}_k \succeq  0, \forall k \in \mathcal{K}$. As a result, $\{\{\hat{\mv{T}}_k\},\hat{\alpha}\}$ is feasible for (SDR3).

Furthermore, it can be shown that for any feasible solution $\{\{\mv{T}_k\},\alpha\}$ to problem (SDR3), 
\begin{align}
	\nonumber
	f_3(\{\hat{\mv{T}}_k\},\hat{\alpha}) &= f_1(\{\tilde{\mv{T}}_k\},\tilde{\mv{R}}_d,\tilde{\alpha}) \\
	&\leq f_1(\{\mv{T}_k\},0,\alpha) = f_3(\{\mv{T}_k\},\alpha),
\end{align}
which shows the optimality of $\{ \{\hat{\mv{T}}_k\},\hat{\alpha} \}$ for (SDR3) and thus completes the proof.

\subsection{Proof of Proposition \ref{T_LOS}}\label{Proof_theorem_LOS}

Let $\{\mv{T}_{k}^{\text{opt}}\}_{k=1}^K$ denote the optimal solution to problem (SDR3). For each $\mv{T}_{k}^{\text{opt}}$, define $\rho_{k} = \text{rank}(\mv{T}_{k}^{\text{opt}}) \geq 1$. Evidently, $\mv{T}_{k}^{\text{opt}}$ can be decomposed as $\mv{T}_{k}^{\text{opt}} = \sum\limits_{\ell=1}^{\rho_{k}} \mv{w}_{k \ell}^{\text{opt}}\left(\mv{w}_{k \ell}^{\text{opt}}\right)^{H}$, and thus the signal power received from $s_k$ at receiver $i$ through channel $\mv{h}_i$ can be written as \cite{karipidis2007far}
\begin{align}
	\label{equ:trQ}
	\nonumber
	& \text{tr}\left(\mv{T}_{k}^{\text{opt}} \mv{h}_{i} \mv{h}_{i}^{H}\right)  \\
	\nonumber
	=&\text{tr}\left[\sum_{\ell=1}^{\rho_{k}} \mv{w}_{k \ell}^{\text{opt}}\left(\mv{w}_{k \ell}^{\text{opt}}\right)^{H} \mv{h}_{i} \mv{h}_{i}^{H}\right] =\sum_{\ell=1}^{\rho_{k}} \text{tr}\left[\mv{w}_{k \ell}^{\text{opt}}\left(\mv{w}_{k \ell}^{\text{opt}}\right)^{H} \mv{h}_{i} \mv{h}_{i}^{H}\right] \\
	\nonumber
	=&\sum_{\ell=1}^{\rho_{k}} \text{tr}\left[\mv{h}_{i}^{H} \mv{w}_{k \ell}^{\text{opt}}\left(\mv{w}_{k \ell}^{\text{opt}}\right)^{H} \mv{h}_{i}\right] =\sum_{\ell=1}^{\rho_{k}}\left|\mv{h}_{i}^{H} \mv{w}_{k \ell}^{\text{opt}}\right|^{2} \\
	=&\sum_{\ell=1}^{\rho_{k}} \left| \mv{v}\left(\phi_{i}\right)^{H} \mv{w}_{k \ell}^{\text{opt}}\right|^{2},
\end{align}
where $\mv{v}\left(\phi_{i}\right) = [1,e^{j \phi_i},...,e^{j (N-1)\phi_i}]^T$. According to the Riesz-Fejer theorem \cite{szeg1939orthogonal}, there exists a vector $\mv{w}_k^{\text{opt}} \in \mathbb{R} \times \mathbb{C}^{N-1}$	that is independent of $\phi_{i}$ such that for all $\phi_{i} = 2 \pi \frac{d}{\lambda} \sin(\theta_i) $,
\begin{align}
	\label{equ:trans}
	\sum_{\ell=1}^{\rho_{k}} \left| \mv{v}\left(\phi_{i}\right)^{H} \mv{w}_{k \ell}^{\text{opt}}\right|^{2}  = | \mv{v}\left(\phi_{i}\right)^{H} \mv{w}_{k }^{\text{opt}}|^{2}
	= \text{tr} (\hat{\mv{T}}_k^{\text{opt}} \mv{h}_{i} \mv{h}_{i}^{H}),
\end{align}
where $\hat{\mv{T}}_k^{\text{opt}} = \mv{w}_{k }^{\text{opt}} (\mv{w}_{k }^{\text{opt}})^H$. Therefore, the new rank-one solution set $\{\hat{\mv{T}}_k^{\text{opt}}\}_{k=1}^K$ will not change the SINR obtained by the original, possibly high-rank solution $\{\mv{T}_{k}^{\text{opt}}\}_{k=1}^K$.

Secondly, we show that the aforementioned reconstructed rank-one solution meets the sum-power constraint at the BS. Given a specific $\phi$ (index $i$ is dropped here for simplicity), we integrate out $\phi$ in (\ref{equ:trans}) at both sides,
\begin{align}
	\frac{1}{2 \pi}\int_{- \pi}^{\pi} \sum_{\ell=1}^{\rho_{k}} \left| \mv{v}\left(\phi\right)^{H} \mv{w}_{k \ell}^{\text{opt}}\right|^{2} d\phi = \frac{1}{2 \pi}\int_{- \pi}^{ \pi} | \mv{v}\left(\phi\right)^{H} \mv{w}_{k }^{\text{opt}}|^{2} d\phi,
\end{align}
which is equivalent to
\begin{align}
	\nonumber
	&\sum_{\ell=1}^{\rho_{k}} (\mv{w}_{k \ell}^{\text{opt}})^H \frac{1}{2 \pi}\int_{-\pi}^{\pi}  \mv{v}(\phi) \mv{v}(\phi)^{H} d\phi \mv{w}_{kl}^{\text{opt}} \\
	= & (\mv{w}_{k}^{\text{opt}})^H \frac{1}{2 \pi}\int_{-\pi}^{\pi}  \mv{v}(\phi) \mv{v}(\phi)^{H} d\phi \mv{w}_{k}^{\text{opt}}.
\end{align}
Denote $\frac{1}{2 \pi}\int_{-\pi}^{ \pi}  \mv{v}(\phi) \mv{v}(\phi)^{H} d\phi$ as $\mv{V} \in \mathbb{C}^{N \times N}$. For the entry $v_{k,m}$ in the $k$-th row and $m$-th column of $\mv{V}$, $k, m \in \{1,2,...,N\}$, we have
\begin{align}
	\nonumber
	v_{k,m} & = \frac{1}{2 \pi} \int_{-\pi}^{\pi} \mv{v}(\phi)_k \mv{v}(\phi)^{H}_m d\phi \\
	\nonumber
	&= \frac{1}{2 \pi} \int_{-\pi}^{\pi} e^{j(k-1)\phi} e^{-j(m-1)\phi} d\phi  \\
	& = \frac{1}{2 \pi} \int_{-\pi}^{\pi} e^{j(k-m)\phi} d\phi = \delta_{km},
\end{align}
where $\delta_{km} = 1$ for $k=m$ and $\delta_{km} = 0$ for $k \ne m$. It thus immediately follows that
\begin{align}
	\nonumber
	& \sum_{\ell=1}^{\rho_{k}} ||\mv{w}_{k \ell}^{\text{opt}}||^2  = ||\mv{w}_{k}^{\text{opt}}||^2 \\
	 \Leftrightarrow & \text{ }
	\text{tr}\left[\sum_{\ell=1}^{\rho_{k}} \mv{w}_{k \ell}^{\text{opt}}\left(\mv{w}_{k \ell}^{\text{opt}}\right)^{H}\right]=\text{tr}\left[\mv{w}_{k}^{\text {opt }}\left(\mv{w}_{k}^{\text{opt}}\right)^{H}\right] \\  \Leftrightarrow & \text{ }
	\text{tr} (\mv{T}_k^{\text{opt}})  = \text{tr} (\hat{\mv{T}}_k^{\text{opt}}) \\
	\Leftrightarrow & \text{ }
	\sum\limits_{k=1}^K \text{tr} (\mv{T}_k^{\text{opt}}) =  \sum\limits_{k=1}^K \text{tr} (\hat{\mv{T}}_k^{\text{opt}}).
\end{align}
Thus, the set of $\{\hat{\mv{T}}_k^{\text{opt}}\}$ also meets the sum power constraints. Furthermore, by letting $\hat{\alpha}^{\text{opt}} = \alpha^{\text{opt}}$, we have
\begin{align}
	\nonumber
	 f_3(\{\mv{T}_k^{\text{opt}}\}, \alpha^{\text{opt}})  & = \sum\limits_{m=1}^{M}\left|\alpha^{\text{opt}} \tilde{\mathcal{P}}_{d}\left(\theta_{m}\right) - \sum\limits_{k=1}^{K} \text{tr}(\mv{T}_k^{\text{opt}} \mv{h}_{m} \mv{h}_{m}^{H})\right|^{2}\\
	 \nonumber
	 & =   \sum\limits_{m=1}^{M}  \left|\hat{\alpha}^{\text{opt}} \tilde{\mathcal{P}}_{d}\left(\theta_{m}\right) - \sum\limits_{k=1}^{K} \text{tr}(\hat{\mv{T}}_k^{\text{opt}}  \mv{h}_{m} \mv{h}_{m}^{H})\right|^{2} \\
	 & = f_3(\{\hat{\mv{T}}_k^{\text{opt}}\}, \hat{\alpha}^{\text{opt}}),
\end{align}
where the second equality follows from (\ref{equ:trQ}) and (\ref{equ:trans}). This shows that the constructed rank-one solution $\{\hat{\mv{T}}_k^{\text{opt}}\}_{k=1}^K$ also has the same objective function value as its original high-rank counterpart $\{\mv{T}_k^{\text{opt}}\}_{k=1}^K$. Combining the above results yields the proof.

%
%
%
%
%
%
%

%
%


\bibliographystyle{ieeetran}
\bibliography{refs}

\end{document}